\documentclass[%
reprint,
 amsmath,amssymb,
 aps,
prb,
superscriptaddress
]{revtex4-2}

\usepackage{graphicx}
\usepackage{dcolumn}
\usepackage{bm}
\usepackage{color}

\def \r {{\bm r}}
\def \R {{\bm R}}

\def \RAi {{\bm R_{{\rm A}i}}}
\def \RBi {{\bm R_{{\rm B}i}}}
\def \RAj {{\bm R_{{\rm A}j}}}

\def \dr {d{\bm r}}
\def \phia {\varphi_{{\rm A}{\bm k}}}
\def \phib {\varphi_{{\rm B}{\bm k}}}

\def \k {{\bm k}}
\def \ek {E}

\def \epm {E_\pm}

\def \ep {\varepsilon}

\def \dks {\Delta}
\def \dksm {\Delta_{\mu}}

\def \dksmn {\Delta_{\mu\nu}}

\def \epx {\varepsilon_x}
\def \epy {\varepsilon_y}

\def \epyy {\varepsilon_{yy}}
\def \epxy {\varepsilon_{xy}}

\def \Hmu {\frac{\partial H_\k}{\partial k_\mu}}
\def \Hnu {\frac{\partial H_\k}{\partial k_\nu}}
\def \Hx {\frac{\partial H_\k}{\partial k_x}}
\def \Hy {\frac{\partial H_\k}{\partial k_y}}

\def \thex {\theta_x}
\def \they {\theta_y}

\def \theyy {\theta_{yy}}
\def \thexy {\theta_{xy}}
\def \them {\theta_\mu}
\def \then {\theta_\nu}

\def \upm {u^\pm_{\k\sigma}}
\def \ump {u^\mp_{\k\sigma}}
\def \upms {u^{\pm *}_{\k\sigma}}
\def \umps {u^{\mp *}_{\k\sigma}}

\def \epm {\varepsilon_\mu}
\def \epn {\varepsilon_\nu}
\def \epmn {\varepsilon_{\mu\nu}}

\def \rsq {\langle y^2 \rangle}

\begin{document}


\title{Topological contribution to magnetism in the Kane-Mele model: \\An explicit wave function approach}

\author{Soshun Ozaki}
 \email{ozaki@hosi.phys.s.u-tokyo.ac.jp}
 \affiliation{Department of Physics, University of Tokyo, Bunkyo, Tokyo 113-0033, Japan}
\author{Masao Ogata}%
 \affiliation{Department of Physics, University of Tokyo, Bunkyo, Tokyo 113-0033, Japan}
 \affiliation{Trans-scale Quantum Science Institute, University of Tokyo, Bunkyo, Tokyo 113-0033, Japan}

\date{\today}

\begin{abstract}
In our previous publication [S.\ Ozaki and M.\ Ogata, Phys. Rev. Research {\bf 3}, 013058], 
the quantization of the orbital-Zeeman (OZ) cross term 
in the magnetic susceptibility, or the cross term of spin Zeeman and orbital effect,
was shown for the Kane-Mele model using the expansion around the Dirac points.
In the present study, we accurately evaluate the orbital, spin-Zeeman, and OZ cross term of the Kane-Mele model
using a recently developed formulation.
This formula is written in terms of the explicit Bloch wave functions, and enables us to evaluate each contribution 
taking account of the integration over the whole Brillouin zone and the summation over all the bands.
As a result, additional contributions such as core-electron diamagnetism are found.
Furthermore, our evaluation confirms the quantization of the OZ cross term 
and reveals its behavior including the metallic case.
The possibility of experimental detection of the quantization is discussed.
\end{abstract}

\maketitle


\section{introduction\label{sec:intro}}
Much research in recent years has focused on topological insulators (TIs).%
\cite{haldane,kane-mele-z2,kane-mele-qsh,bernevig-prl,bernevig-sci,yang-chang, murakami06,konig,roth,brune,
knez, knez12, fu07, hsieh}
TIs show anomalous phenomena such as electric conduction on 
sample surfaces, and the search for candidate materials is one of the most 
important problems in this field.
In particular, the novel phenomena such as spin Hall effect and the conducting edge state robust against 
nonmagnetic impurities are expected in two-dimensional (2D) TIs.
However, only a few materials have been confirmed to be 2D TIs.\cite{konig,roth,brune,knez,knez12}
So far, the confirmation of TI is achieved by finding the edge state in
angle-resolved photoemission spectroscopy and in the transport experiments,
which are both by the methods to find the anomalous edge states.
Therefore, it is desirable to develop some alternative methods 
that detect the topological nature of a material through some observation of 
bulk physical quantities.
One example for such quantities is the magnetic susceptibility.

Usually, the magnetic susceptibility in itinerant systems without two-body interactions 
is discussed in terms of the spin Zeeman effect and the orbital effect individually.
Generally, however, in the system with spin-orbit interaction (SOI) the cross term of the two exists,
which we call ``orbital-Zeeman (OZ) cross term'' in the following.
Although the OZ cross term were discussed for some situations 
\cite{murakami06,nakai-nomura,loss2015, koshino2016, suzuura-ando,nomura2019, aftab,ozaki-ogata},
further research on it is desired.

Nakai and Nomura \cite{nakai-nomura} reported that the OZ cross term $\chi_\mathrm{OZ}$ in the 
Bernevig-Hughes-Zhang model \cite{bernevig-sci}
is proportional to the spin Chern number $\mathrm{Ch}_{\mathrm{s},l}$,
the topological invariant for spin-conserving 2D TIs.
Recently, we clarified that the coefficient 
of $\mathrm{Ch}_{\mathrm{s},l}$ is generally given by
the universal value $\chi_u=4|e|\mu_B/h$,
and confirmed the quantization in the Kane-Mele model \cite{kane-mele-z2, kane-mele-qsh} by using the ${\bm k} \cdot {\bm p}$ 
approximation explicitly \cite{ozaki-ogata}.
This quantization is associated with the Berry curvature, and physically, 
it originates from the edge currents characteristic of the 2D TIs \cite{ozaki-ogata}.
Based on these results, the magnetic susceptibility is expected to be used for 
the detection of the change in the topological invariant for 2D TIs.
However, the evaluation for the Kane-Mele model in the previous paper was based on the effective Hamiltonian 
in the vicinities of K and K$'$ points of graphene, where the massless Dirac electron 
system is realized.
Therefore, the contributions from the distant ${\bm k}$ region from the K or K$'$ points 
or from the bands other than the two bands forming the Dirac dispersion were not evaluated precisely.
Besides, the previous publication does not include some additional contributions,
such as core-electron diamagnetism, the newly found Fermi surface term,
and the correction term from the occupied states \cite{ogata-fukuyama, ogata2,ogata3}.
To compare the theory with experiments, it is desirable to be able to evaluate all these contributions.

%
%
Therefore, in this paper, we clarify all the contributions in magnetic susceptibility including the spin Zeeman,
orbital, and OZ cross term in the Kane-Mele model, taking care of the contributions from 
the whole Brillouin zone.
We will show that the OZ cross term has a reasonable magnitude compared with the other contributions and confirm the 
quantized jump at the topological phase transition.
We also study the chemical potential dependence of each contribution and find that the OZ cross term also has a 
contribution at the van Hove singularity irrespective of whether the system is topological or not.

Here, it should be noted that the Kane-Mele model is based on the tight-binding model of graphene.
The effect of the magnetic field is often introduced as the Peierls phase of the transfer integral 
in the tight-binding model.
However, it is known that the Peierls phase is not enough to describe the effect of the magnetic field \cite{lowdin, matsuura2016}.
To obtain the whole magnetic susceptibility, it is necessary to use the contiuum Hamiltonian with the SOI.
The formulation of the orbital magnetism in such a case is not so simple 
owing to the complicated interband effects of the magnetic field,
and a lot of efforts were dedicated \cite{landau, peierls,fuku1970,hebborn1960,blount,hebborn1964,fukuyama}.
It is notable that Fukuyama developed a one-line formula for the orbital magnetic susceptibility,
which is written in terms of Green's functions \cite{fukuyama}.
(Some recent publications \cite{gomez, raoux-piechon} proposed similar formulae.)
Fukuyama's formula was reformulated in terms of Bloch wave functions 
with infinite summations over the band indices taken analytically \cite{ogata-fukuyama, ogata2017},
and the resultant formula properly includes the Landau-Peierls (dia)magnetism and the contribution from occupied states,
and some additional contributions.
The total magnetic susceptibility is given by
\begin{align}
\chi_{\rm total}=\chi_{\rm LP} + \chi_{\rm inter} + \chi_{\rm FS} + \chi_{\rm FS-P}
+\chi_{\rm occ1} + \chi_{\rm occ2},
	\label{eq:sixcontrib}
\end{align}
where $\chi_{\rm LP}$, $\chi_{\rm inter}$, $\chi_{\rm FS}$, and $\chi_{\rm occ1}$ represent
Landau-Peierls, interband, Fermi surface, and occupied state contributions, respectively.
$\chi_{\rm FS-P}$ is also the Fermi surface contribution, which contains Pauli paramagnetism.
$\chi_{\rm occ2}$ is a Berry curvature-related term.
The explicit expressions for these contributions are shown in Ref.~\cite{ogata2017}.
To apply the above formula to the tight-binding model of Kane-Mele, we use the method of linear combination
of atomic orbitals (LCAOs) for the $p_z$ orbitals on carbon atoms as in the case of simple graphene \cite{ogata3}.
Then, we evaluate Eq.~\eqref{eq:sixcontrib} using the obtained wave functions and study various contributions in the orbital term,
spin Zeeman term, and OZ cross term.

This paper is organized as follows.
In Sec.~\ref{sec:model}, we first review the fundamental properties of the Kane-Mele model 
such as the energy dispersion and topological phase diagram.
Then, we derive the continuum-space Bloch wave function assuming that it consists of $2p_z$ atomic orbitals.
In Sec.~\ref{sec:analytic}, we apply the general formula for the magnetic susceptibility to the 
Kane-Mele model and derive analytic expressions for each contribution.
In Sec.~\ref{sec:numerical}, we numerically evaluate the results obtained in the previous section and 
discuss the experimental observability of each term, especially the OZ cross term.
Section~\ref{sec:summary} is devoted to summary.

\section{explicit wave functions for the Kane-Mele model\label{sec:model}}

\begin{figure}[h]
	\centering\includegraphics[width=5cm]{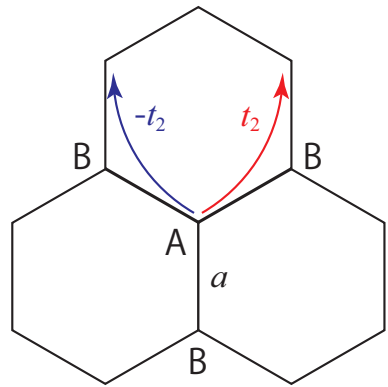}
	\caption{\label{honeycomb} Honeycomb lattice for the Kane-Mele model. A and B represent the sublattices and 
	$a$ is the distance between the adjacent two sites.
	The arrows show the hopping between the next-nearest-neighbors due to SOI.
	The signs of this hopping depends on the path: It is $+1(-1)$ if the electron makes a left (right) turn to 
	propagate to a next-nearest site.}
\end{figure}

We study the energy dispersion and write down the explicit wave functions for the Kane-Mele model.
The tight-binding Hamiltonian is given by \cite{kane-mele-qsh,kane-mele-z2}
\begin{align}
	H=-&\sum_{\langle i,j \rangle \alpha}t c_{i\alpha}^\dagger c_{j\alpha} 
	+\Delta_0 \left(\sum _{i \in A, \alpha} c_{i\alpha}^\dagger c_{i\alpha} 
	- \sum_{i\in B,\alpha}c_{i\alpha}^\dagger c_{i\alpha} \right ) \nonumber \\
  &+\sum_{\langle\langle i,j \rangle\rangle, \alpha,\beta}  i t_2 \nu_{ij}
c_{i\alpha}^\dagger \sigma_{\alpha \beta}^z  c_{j\beta},\label{tbham}
\end{align}
where $c^\dagger_{i\alpha}$ creates an electron with spin $\alpha$ at
site $i$, and $\langle i,j \rangle$ ($\langle \langle i,j \rangle \rangle$)
run over all the nearest- (next-nearest-) neighbor sites of a two-dimensional honeycomb lattice
defined in Fig.~\ref{honeycomb}.
The first term represents usual nearest-neighbor hoppings
with transfer integral $t$.
The second term represents a staggered on-site potential,
$+\Delta_0$ for the sites in the A sublattice and $-\Delta_0$ for those in the B sublattice.
The summation $\sum_{i \in {\rm A}}$ ($\sum_{i\in {\rm B}}$) means the sum over the 
sites in the A (B) sublattice. 
The last term represents the hopping between the next-nearest-neighbor sites due to SOI,
where $\sigma^z_{\alpha\beta}$ is the $(\alpha\beta)$ component of the spin operator in the $z$ direction, 
and $\nu_{ij}=-\nu_{ji}=+1 \, (-1)$ if the electron makes a left (right) turn to propagate 
to a next-nearest site [see Fig.~\ref{honeycomb}].
Only the $\sigma^z$ component of SOI appears,~\cite{kane-mele-qsh,ezawanjp,ezawaepj} 
whose microscopic derivation using the LACOs is shown in Appendix A.
This model is known as the model for silicene
\cite{gvv,ljy,ezawanjp,ezawaepj,lfy},
in which we can control $\Delta_0$ by changing the electric field applied perpendicularly
to the layer owing to the buckled structure of silicene.

As discussed in Sec.~\ref{sec:intro}, to include all the effect of a magnetic field correctly,
we consider the Hamiltonian in the continuum space,
\begin{align}
	H^{\rm full}=&\frac{1}{2m}({\bm p}-e{\bm A})^2+V({\bm r})+\frac{\hbar^2}{8m^2c^2}\nabla^2 V \nonumber \\
	&+\frac{\hbar}{4m^2c^2}\boldsymbol \sigma \cdot \nabla V \times ({\bm p}-e{\bm A})
	-\frac{e\hbar}{2m}\boldsymbol \sigma \cdot {\bm B},
	\label{eq:hamfull}
\end{align}
where $e(<0)$ is the electron charge, ${\bm A}({\bm r})$ is a vector potential 
(${\bm B}={\rm rot}{\bm A}$).
$V({\bm r})$ is the periodic potential that represents the honeycomb lattice 
and the fourth term represents the SOI
derived from the relativistic Dirac equation.
We have set the \textit{g} factor in the last term (spin Zeeman term) to be $g=2$ neglecting
the QED corrections.
When we consider a tight-binding model under a magnetic field,
a renormalization of the \textit{g} factor generally occurs owing to the virtual interband process \cite{roth59,sczhang},
and the effective $g$ factor deviates from $g=2$ in the focused bands.
However, since the Hamiltonian Eq.~\eqref{eq:hamfull} contains all the bands,
the Zeeman term in Eq.~\eqref{eq:hamfull} should have the bare $g$ factor, $g=2$.
At the end of calculation including the interband processes, the effective $g$ factor should naturally appear.

We apply Eq.~(\ref{eq:hamfull}) to the Kane-Mele model.
In Eq.~\eqref{eq:hamfull}, $V({\bm r})$ is chosen to be the periodic potential formed by 
the atoms on the honeycomb lattice,
\begin{equation}
	V({\bm r})=\sum_{i\in i{\rm A}}V_{\rm A}({\bm r}-{\bm R}_{{\rm A}i})+\sum_{i\in {\rm B}}V_{\rm B}({\bm r}-{\bm R}_{{\rm B}i}),
\end{equation}
where ${\bm R}_{{\rm A}i}$ (${\bm R}_{{\rm B}i}$) represents the position of the site in the A (B) sublattice in the 
$i$th unit cell.
For the continuum Hamiltonian Eq.~\eqref{eq:hamfull}, we write down the Bloch wave functions in terms of 
LCAOs of the A and B sublattice \cite{ogata3},
assuming that the wave functions near the Fermi level consist of 2$p_z$ orbitals.
Then, the wave functions are expressed as the linear combinations of the orthogonal wave functions 
localized at a site ${\bm r}={\bm R_i}$.
(The other bands are treated later.)
Using the wave function for $2p_z$ orbital,
\begin{align}
	\phi_{2p_z}({\bm r})=\frac{1}{\sqrt{24}(a_{\rm B}^*)^{5/2}} \sqrt{\frac{3}{4\pi}}z e^{-r/2a_{\rm B}^*}
\end{align} 
with $a_{\rm B}^*$ being the renormalized Bohr radius, the orthogonal localized basis is given by \cite{lowdin, ogata2,ogata3}
\begin{align}
  \Phi(\r -\R_i) = \phi_{2p_z}(\r-\R_i) - \sum_{j:\rm{n.n.}}\frac{s}{2}\phi_{2p_z}(\r-\R_j). \label{eq:orthogonal}
\end{align}
Here, the renormalized Bohr radius is $a_{\rm B}^*=a_{\rm B}/Z_{\rm eff}$,
where $a_{\rm B}=\hbar^2/me^2$ is the Bohr radius and $Z_{\rm eff}=3.25$\cite{slater,ogata3} is the effective charge of carbon atoms.
In Eq.~\eqref{eq:orthogonal}, $j$ summation is taken over the 
nearest-neighbor (n.n.) sites of ${\bm R}_i$, and $s$ is the overlap integral between the adjacent sites,
\begin{align}
  s=\int \phi_{2p_z}^*(\r-\R_i) \phi_{2p_z}(\r-\R_j) \dr.
\end{align}
The orthogonality of $\Phi({\bm r}-{\bm R}_i)$ is maintained up to the first order 
with respect to $s$.
Note that $s$ is independent of the direction $\R=\R_j-\R_i$ 
since the $p_z$ orbital is isotropic in the $xy$-plane.

Next, we perform a Fourier transform and obtain the basis
\begin{align}
\varphi_{{\rm A}\k}(\r)=\frac{1}{\sqrt{N}}\sum_{\RAi} e^{-i\k(\r-\RAi)}\Phi(\r-\RAi),\label{eq:phia}
\end{align}
and
\begin{align}
\varphi_{{\rm B}\k}(\r)=\frac{1}{\sqrt{N}}\sum_{\RBi} e^{-i\k(\r-\RBi)}\Phi(\r-\RBi),\label{eq:phib}
\end{align}
where $N$ is the total number of sites on each sublattice.
The periodic part of the Bloch wave function $u_{l\k\sigma}({\bm r})$ is determined by the eigenvalue equation,
\begin{equation}
	H_{\k\sigma} u_{l\k\sigma} = E_{l\k\sigma}u_{l\k\sigma}, \qquad H_{\k\sigma}=e^{-i\k{\bm r}} H^{\rm full} e^{i\k{\bm r}}, \label{eq:eigen}
\end{equation}
where $l$ and $\sigma(=\pm 1)$ represent the band index and the eigenvalue of the $z$-component of spin of an electron: $\sigma=1$ 
for spin-up and $\sigma=-1$ for spin-down.
We denote the two energy dispersion and the two eigenfunctions near the Fermi level as 
$E_{\bm k \sigma}^\pm$ and $u^\pm_{{\bm k}\sigma}({\bm r})$, respectively.
To determine $E_{\bm k \sigma}^\pm$ and $u^\pm_{{\bm k}\sigma}({\bm r})$, we calculate the matrix elements of the Hamiltonian $H_{\bm k \sigma}$ 
in terms of the obtained basis,
\begin{equation}
	h_{nm\sigma}({\bm k})=\int \varphi^*_{n\k}({\bm r})H_{\k\sigma} \varphi_{m\k}({\bm r}) d{\bm r}, \label{eq:hammatrix}
\end{equation}
with $n,m={\rm A, or\,B}$.
They become
\begin{eqnarray}
	&h_{\rm AA\sigma}=\Delta_{\k\sigma} + E_0, \\
	&h_{\rm BB\sigma}=-\Delta_{\k\sigma} + E_0, \\
	&h_{\rm AB\sigma}=h^*_{\rm BA\sigma}=-t\gamma_\k,
\end{eqnarray}
where $E_0$ is the energy constant \cite{ogata3},
\begin{align}
	\Delta_{\k \sigma} = \Delta_0  +4 \sigma t_2 \sin\frac{\sqrt{3}}{2} k_x a
		\left(\cos\frac{3}{2}k_y a - \cos\frac{\sqrt{3}}{2} k_x a \right), \label{eq:delta}
\end{align}
and,
\begin{align}
	\gamma_\k= e^{-ik_y a} + e^{i(\frac{\sqrt{3}}{2}k_x + \frac{1}{2}k_y)a}
		+e^{i(-\frac{\sqrt{3}}{2}k_x + \frac{1}{2}k_y)a},
\end{align}
with $a$ being the distance between the nearest-neighbor sites.
Hereafter, we set $E_0=0$ without loss of generality.
$\Delta_0$ corresponds to the staggered on-site potential in Eq.~(\ref{tbham}) 
and the nearest-neighbor hopping $t$ can be expressed as a kind of overlap integral \cite{ogata3}.
As shown in Appendix \ref{sec:spinorbit}, the SOI
in Eq.~(\ref{eq:hamfull}) does not give contributions to the nearest-neighbor hopping,
but it leads to a next-nearest-neighbor hopping $t_2$ in Eq.~(\ref{tbham}),
which is also expressed by a kind of the overlap integrals.
In Appendix~\ref{sec:spinorbit},
it is also shown that the next-nearest-neighber hopping have only 
the $z$ component of spin $\sigma_{\sigma\sigma'}^z$;
therefore, the spin is conserved.

\begin{figure}[h]
	\centering\includegraphics[width=\linewidth]{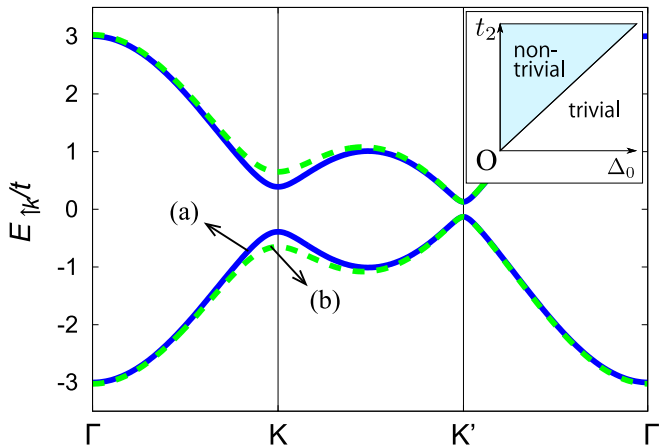}
	\caption{\label{band} Energy dispersion of Eq.~(2) for $\sigma=1$ (up spin)
	along the path $\Gamma \rightarrow$ K $\rightarrow$ K $'\rightarrow \Gamma$
	for two typical choices of parameters:
	(a) solid line, $\Delta_0/t=1/4,t_2/t=\sqrt{3}/72$ (topologically trivial)
	and (b) dashed line, $\Delta_0/t=1/4,t_2/t=\sqrt{3}/24$ (topologically nontrivial).
		The energy dispersion for $\sigma=-1$ (down spin) is obtained by exchanging K for K$'$ points.
	Inset:
	Phase diagram of the present model.
	This figure is taken from Ref.\cite{ozaki-ogata}.	}
\end{figure}

%
By diagonalizing the Hamiltonian $h_{nm\sigma}({\bm k})$, we obtain the energy dispersion,
\begin{equation}
E^\pm_{\k\sigma} = \pm \sqrt{\Delta_{\k \sigma}^2 + \varepsilon_\k^2}, \label{eq:enedisp}
\end{equation}
where $\varepsilon_{\bm k} := t|\gamma_\k|$.
At ${\rm K}=(4\pi/3\sqrt{3}a,0)$ and ${\rm K}'=(-4\pi/3\sqrt{3}a,0)$ in the Brillouin zone, $\gamma_{\bm k}$ vanishes.
Figure \ref{band} shows the energy dispersions Eq.~(\ref{eq:enedisp}) with $\sigma=+1$ (up spin)\cite{ozaki-ogata}.
The solid and the dashed lines in Fig.~\ref{band} are for the cases of
$\Delta_0/t=1/4$, $t_2/t=\sqrt{3}/72$ (topologically trivial) and
$\Delta_0/t=1/4$, $t_2/t=\sqrt{3}/24$ (topologically nontrivial), respectively.
The dispersions are similar to those of graphene but gaps open at K and K$'$ points.
The magnitudes of the gaps at K and K$'$ points are given by
\begin{align}
	2\left| \Delta_0 + 3\sqrt{3} \sigma t_2 \right|,  \quad 
	2\left| \Delta_0 - 3\sqrt{3} \sigma  t_2 \right|, \label{gap1}
\end{align}
respectively.
Owing to the SOI, the energy dispersions for spin-up and spin-down 
are not the same.
The energy dispersion for $\sigma=-1$ (down spin) is obtained by exchanging K for K$'$ points in Fig.~\ref{band}.
As shown in the inset of Fig.~\ref{band} \cite{kane-mele-qsh,kane-mele-z2,ezawanjp, ezawaepj},
the system is topologically trivial for $|\Delta_0|>3\sqrt{3}|t_2|$
while the system is topologically nontrivial for $|\Delta_0|<3\sqrt{3}|t_2|$.
As usual, the gap closes at the nontrivial-trivial critical points, $\Delta_0=\pm3\sqrt{3}t_2$.

The two eigenfunctions, $u_{\l\sigma_{{\bm k}\sigma}}^\pm({\bm r})$, are obtained as
\begin{align}
u_{\k\sigma}^+ (\r) =e^{\frac{i}{2}\theta_\k} \cos \eta_{\k\sigma} \varphi_{{\rm A}\k}(\r)
- e^{-\frac{i}{2}\theta_\k} \sin \eta_{\k\sigma} \varphi_{{\rm B}\k}(\r),
\label{eq:wfplus}
\end{align}
and
\begin{align}
u_{\k\sigma}^- (\r) =e^{\frac{i}{2}\theta_\k} \sin \eta_{\k\sigma} \varphi_{{\rm A}\k}(\r)
+ e^{-\frac{i}{2}\theta_\k} \cos \eta_{\k\sigma} \varphi_{{\rm B}\k}(\r),
\label{eq:wfminus}
\end{align}
where
\begin{align}
&e^{i\theta_\k}=\frac{\gamma_k}{|\gamma_k|}, \label{eq:deftheta}\\
&\cos \eta_{\k\sigma}=\sqrt{\frac{1}{2} \left( 1+ \frac{\Delta_{\k\sigma}} {|E^\pm_{\k\sigma}|} \right)},\label{eq:defeta1}\\
&\sin \eta_{\k\sigma}=\sqrt{\frac{1}{2} \left( 1- \frac{\Delta_{\k\sigma}} {|E^\pm_{\k\sigma}|} \right)} \label{eq:defeta2}.
\end{align}
If we set $\eta_{{\bm k}\sigma}=\frac{\pi}{4}$, these eigenfunctions coincide with those for 
(massless) graphene \cite{ogata3}.

Note that the Hamiltonian Eq.~\eqref{eq:hamfull} contains all the bands.
In the following, the eigenenergies and eigenfunctions of all the other 
bands are denoted as $E_{l'\k\sigma}$ and $u_{l'\k\sigma}({\bm r})$ with 
$l'\neq \pm$.
As we will show later, they are used in the interband contribution of 
magnetic susceptibility $\chi_{\rm inter}$, but 
the explicit forms of $E_{l'\k\sigma}$ and $u_{l'\k\sigma}({\bm r})$ 
are not necessary.

\section{Magnetic susceptibility\label{sec:analytic}}

Generally, the magnetic susceptibility consists of six contributions shown in Eq.~(\ref{eq:sixcontrib})\cite{ogata-fukuyama,ogata2017}.
Using the eigenenergies and eigenfunctions mentioned above, each term in Eq.~(\ref{eq:sixcontrib}) becomes
\begin{widetext}
\begin{align}
  \chi_{\rm LP} &= \frac{e^2}{12\hbar^2} \sum_{\pm,\k \sigma} f'(E^\pm_{\k\sigma})
  \left[ \frac{\partial^2 E^\pm_{\k\sigma}}{\partial k_x^2} 
	\frac{\partial^2 E^\pm_{\k\sigma}}{\partial k_y^2}
   -\left(\frac{\partial^2 E^\pm_{\k\sigma}}{\partial k_x \partial k_y}\right)^2 \right]  \\
   \chi_{\rm inter}&= - 2 \sum_\pm \sum_{l'\neq\pm,\mp,}\sum_{\k,\sigma} 
	 \frac{f(E^\pm_{\k\sigma})}{E^\pm_{\k\sigma} - E_{l'\k\sigma}}
	  |M_{\pm l'\sigma}|^2  \label{eq:chiinter}\\
   \chi_{\rm FS} &=  \frac{e^2}{2\hbar^2} {\rm Re} \sum_{\pm,\k,\sigma} f'(E^\pm_{\k\sigma}) 
   \left [ \left\{ \frac{\partial E^\pm_{\k\sigma}}{\partial k_x} \int \frac{\partial \upms}{\partial k_y} 
	 \left ( \Hx + \frac{\partial E^\pm_{\k\sigma}}{\partial k_x} \right )
   \frac{\partial \upm}{\partial k_y }\dr  \right. \right. \nonumber \\
   &-\left. \left.  \frac{\partial E^\pm_{\k\sigma}}{\partial k_x} \int \frac{\partial \upms}{\partial k_x} 
	 \left ( \Hy + \frac{\partial E^\pm_{\k\sigma}}{\partial k_y} \right ) 
   \frac{\partial \upm}{\partial k_y }\dr
	 +  (x \leftrightarrow y) \right \} \right . 
   - \left . \left \{ \frac{i\hbar^2}{m} \frac{\partial E^\pm_{\k\sigma}}{\partial k_x} 
	 \int \upms \sigma_z \frac{\partial \upm}{\partial k_y} \dr
   - (x \leftrightarrow y) \right \}\right ]  \\
   \chi_{\rm FS-P} &= -\sum_{\pm,\k,\sigma} f'(E^\pm_{\k\sigma}) |M_{\pm\pm\sigma}|^2  \\
   \chi_{\rm occ1} &= -\frac{e^2}{4\hbar^2} \sum_{\pm,\k,\sigma} f(E^\pm_{\k\sigma})
   \left [ \frac{\partial^2 E^\pm_{\k\sigma}}{\partial k_x \partial k_y} 
   \int \frac{\partial \upms}{\partial k_x} \frac{\partial \upm}{\partial k_y} \dr
   + \left(\frac{\hbar^2}{m} - \frac{\partial^2 E^\pm_{\k\sigma}}{\partial k_x^2}\right)
   \int \frac{\partial \upms}{\partial k_y} \frac{\partial \upm}{\partial k_y} \dr
   \right ] + (x \leftrightarrow y)  \\
   \chi_{\rm occ2} &= -\frac{e}{\hbar} {\rm Re} \sum_{\pm,\k,\sigma} 
	 f(E^\pm_{\k\sigma}) M_{\pm\pm\sigma} \Omega_{\pm\sigma}, \label{eq:occ2}
\end{align}
\end{widetext}
where $M_{l l'\sigma}$ is the magnetic moment defined by
\begin{align}
  M_{ll'\sigma} =&-\frac{ie}{2\hbar} \left\{ \int \frac{\partial u_{l\k\sigma}^*}{\partial k_x} 
	\left(\Hy+\frac{\partial E_{l\k\sigma}}{\partial k_y}\right) u_{l'\k\sigma} \dr  \right. \nonumber \\
	&-  \left. \int \frac{\partial u_{l\k\sigma}^*}{\partial k_y} \left(\Hx+\frac{\partial E_{l\k\sigma}}{\partial k_x} 
	\right) u_{l'\k\sigma} \dr \right\}  \nonumber \\
	&+\frac{e\hbar}{2m} \int u_{l\k\sigma}^* \sigma_z u_{l'\k\sigma} d{\bm r}
	\label{eq:magmoment},
\end{align}
and 
$\Omega_{\pm \sigma}$ is the $z$-component of the Berry curvature
\begin{equation}
	\Omega_{\pm \sigma}= 
	i  \int d{\bm r} \left( \frac{\partial u^{\pm *} _{\k\sigma}}{\partial k_x} 
	\frac{\partial u^\pm_{\k\sigma}}{\partial k_y}
	- \frac{\partial u^{\pm *} _{\k\sigma}}{\partial k_y} 
	\frac{\partial u^\pm_{\k\sigma}}{\partial k_x}
	\right).
\end{equation}
In principle, there are contributions of core level electrons (i.e., in the 1$s$ orbital etc.)
in $\chi_{\rm occ1}$ and $\chi_{\rm occ2}$, which we do not consider in the following.

Various integrals appearing in the above equations are calculated by using the Bloch 
wave functions in Eqs.~(\ref{eq:wfplus}) and (\ref{eq:wfminus})  
up to the first order with respect to 
the \lq\lq overlap integrals'' $s$, $t$, and $t_2$, whose
integrands contain the overlap of atomic orbitals 
$\phi_{2p_z}^*(\r-\R_j)\phi_{2p_z}(\r-\R_i)$ with $\R_i$ and $\R_j$ being the 
nearest-neighbor sites or next-nearest-neighbor sites.
The obtained integrals are shown in Appendix B. 
In the following, we write 
$E_{\k\sigma} := \sqrt{\Delta_{\k \sigma}^2 + \varepsilon_\k^2}$ (i.e., $E^\pm_{{\bm k}\sigma}=\pm E_{\k\sigma}$).
Furthermore, to simplify the expressions 
we abbreviate $\varepsilon_\k$, $\theta_\k$, $E_{\k\sigma}$, 
$\Delta_{\k\sigma}$, and $\eta_{\k\sigma}$ as $\varepsilon$, $\theta$, $E$, 
$\Delta$, and $\eta$, respectively, as far as they do not cause ambiguity.
We also use the abbreviations,
\begin{align}
	\epm&=\frac{\partial \ep}{\partial k_\mu}, \quad
	\theta_\mu= \frac{\partial \theta}{\partial k_\mu}, \quad
	E_\mu = \frac{\partial E}{\partial k_\mu}, \quad
	\dksm=\frac{\partial \dks}{\partial k_\mu}, \nonumber \\
	\eta_\mu &= \frac{\partial \eta}{\partial k_\mu},\quad
	\epmn=\frac{\partial^2 \ep}{\partial k_\mu \partial k_\nu}, \quad
	\theta_{\mu\nu}=\frac{\partial^2 \theta}{\partial k_\mu \partial k_\nu}, \nonumber \\
	E_{\mu\nu}&=\frac{\partial^2 E}{\partial k_\mu \partial k_\nu}, \quad
	\dksmn=\frac{\partial^2 \dks}{\partial k_\mu \partial k_\nu}, \quad
	\eta_{\mu\nu}=\frac{\partial^2 \eta}{\partial k_\mu \partial k_\nu}, 
\end{align}
for $\mu,\nu=x\,{\rm or}\, y$. 
For example, using the formula (F3) in Appendix B, we obtain 
the Berry curvature as
\begin{equation}
\Omega_{\pm \sigma} = 
\mp \frac{\varepsilon}{E}
(\theta_x \eta_y - \theta_y \eta_x) + O(s^2),
\end{equation}
On the other hand, using the formulae (F1) and (F7) in Appendix B, 
the diagonal matrix element of magnetic moment $M_{\pm\pm\sigma}$ becomes 
\begin{equation}
M_{\pm\pm \sigma} = \frac{e}{2\hbar} (\Delta_x \theta_y - \Delta_y \theta_x) + 
\frac{e\hbar}{2m} \sigma +O(s^2).
\end{equation}
Here we have used a relation
\begin{equation}
\frac{\Delta}{E} E_\mu - 2\varepsilon \eta_\mu  = \Delta_\mu,
\label{eq:SpRel1text}
\end{equation}
which are shown in Appendix C. Other useful relations are also shown in Appendix C. 

From the explicit forms of $E, \Omega_{\pm\sigma}$, and $M_{\pm\pm\sigma}$, 
it is straightforward to write down $\chi_{\rm LP}, \chi_{{\rm FS-P}}$, and 
$\chi_{\rm occ2}$. 
$\chi_{{\rm FS}}$ and $\chi_{\rm occ1}$ are shown in Appendix D, where we have used the integral formulae in Appendix~B. 
$\chi_{\rm inter}$ contains the summation over $E_{l'\k\sigma}$ and $u_{l'\k\sigma}$,
which are the other energy dispersions and wave functions than 
for the two bands forming the Dirac dispersion.
We can calculate the summation over $l'$ without using
the explicit expression of $u_{l'\k\sigma}$ by making use of 
the completeness condition,
\begin{align}
	&u^+_{\k\sigma}({\bm r})u_{\k\sigma}^{+*}({\bm r'}) + u^{-}_{\k\sigma}({\bm r})u_{\k\sigma}^{-*}({\bm r'}) 
	+ \sum_{l'\neq \pm,\sigma}u_{l'\k\sigma}({\bm r})u_{l'\k\sigma}^*({\bm r'}) \nonumber \\
	&\quad=\delta({\bm r}-{\bm r'}).
\end{align}
The details of calculations are shown in Appendix E. 

%
%
$\chi_{\rm total}$ [Eq.~\eqref{eq:sixcontrib}] is calculated in Appendix D, which
is classified into a few groups as follows:
\begin{equation}
	\chi_{\rm total}=\chi_{\rm LP} + \chi_{\rm Pauli} + \chi_{\rm OZ} 
	+ \chi_{\rm atomic} + \chi_1+ \chi_2,
	\label{eq:TotalChi1}
\end{equation}
with 
\begin{widetext}
\begin{align}
  \chi_{\rm LP} &= \frac{e^2}{12\hbar^2} \sum_{\pm,\k \sigma} f'(\pm E)
  \left[ E_{xx} E_{yy} - E_{xy}^2 \right],  \label{eq:chi_LP} \\
  \chi_{\rm Pauli} &= -\frac{e^2\hbar^2}{4m^2} \sum_{\pm,\k \sigma} f'(\pm E), \label{eq:chi_Pauli} \\
   \chi_{\rm OZ} &= \frac{e^2}{m} \sum_{\pm,\k,\sigma} f'(\pm E) 
   \sigma \varepsilon(\eta_x \theta_y -\eta_y \theta_x)  
-\frac{e^2}{m} \sum_{\pm,\k,\sigma} f(\pm E) \sigma \Omega_{\pm \sigma} +O(s^2), \label{eq:chi_OZ} \\
\chi_{\rm atomic} &= - \frac{e^2}{4m} \sum_{\pm,\k,\sigma} f(\pm E) \langle x^2+y^2 \rangle
=-\frac{3e^2 a_{\rm B}^{*2}}{m}n(\mu)+O(s^2)
\label{eq:chi_atomic} \\
  \chi_1 &= \frac{e^2}{\hbar^2} \sum_{\pm,\k,\sigma} f(\pm\ek) \biggl[ \pm \frac{\eta_x^2}{2\ek}(\varepsilon\epyy +\Delta\Delta_{yy}) 
    \mp \frac{\eta_x \eta_y}{2\ek} (\varepsilon \epxy + \Delta\Delta_{xy}) 
    \mp \frac{\ep\Delta}{4\ek}(\eta_x \thex\theyy - \eta_x \they\thexy) \nonumber \\
  & \pm \frac{1}{8}(E_x \thex\theyy - E_x \they\thexy) 
  \mp \frac{\thex^2}{8\ek}(\Delta\Delta_{yy} +2 \varepsilon \Delta_y\eta_y) 
  \pm \frac{\thex \they}{8\ek} (\Delta\Delta_{xy} +2 \varepsilon \Delta_x \eta_y) 
  \biggr] + (x\leftrightarrow y) \nonumber \\
 &+\frac{e^2}{\hbar^2} \sum_{\pm\k\sigma} f'(\pm E) 
	\biggl[ \frac{\varepsilon^2 E_x}{8E} (\they\thexy - \thex\theyy) 
	+\frac{\Delta E_x}{8E}(\Delta_x \they^2 - \Delta_y\thex\they) \nonumber \\
	&+ \frac{E_x}{4E} \left\{ \eta_y (\Delta \varepsilon_{xy}-\varepsilon \Delta_{xy}) - \eta_x (\Delta\varepsilon_{yy}-\varepsilon \Delta_{yy})\right\} 
	-\frac{1}{4} \left( \theta_x^2 \Delta_y^2 - \theta_x \theta_y \Delta_x \Delta_y \right) \biggr] 
	+(x\leftrightarrow y)  + O(s^2),
  \label{eq:chi_1} \\
  \chi_{2} &= \frac{e^2}{\hbar^2} \sum_{\pm,\k,\sigma} f(\pm\ek) 
  \biggl[  \mp \frac{\rsq}{4\ek}
	(- a^2\varepsilon^2 + \Delta \Delta_{xx} +\Delta \Delta_{yy}) 
  \mp \frac{\hbar^2}{4m} \left(\frac{a^2 s\varepsilon^2}{2Et} 
  - \frac{\varepsilon^2}{Et} \langle x^2+y^2 \rangle_R \right)\biggr] +O(s^2). 
  \label{eq:chi_2}
\end{align}
\end{widetext}
The \lq\lq expectation values'' $\langle \cdots \rangle$ and $\langle \cdots \rangle_R$ are defined by
\begin{align}
	\langle x^2+y^2 \rangle = \int \Phi^* (\r) (x^2+y^2) \Phi(\r) \dr,
\end{align}
and
\begin{align}
	\langle x^2+y^2 \rangle _R = \int \Phi^*(\r-\R) (x^2+y^2) \Phi(\r) \dr,
\end{align}
with ${\bm R}$ being one of the vectors that point adjacent sites.
Note that the value $\langle x^2+y^2 \rangle _R$ depends on $R=|{\bm R}|$,
and does not depend on the direction of ${\bm R}$.
The present result is consistent with the previous one for graphene,\cite{ogata3}
but there appear several new contributions 
due to the presence of the staggered on-site potential $\Delta_0$ and 
SOI $t_2$.

$\chi_{\rm Pauli}$ is the Pauli paramagnetism and $\chi_{\rm OZ}$ is the OZ cross term obtained in \cite{ozaki-ogata}.
$\chi_{\rm atomic}$ represents the contributions from the occupied states in the partially filled
2$p_z$-band, which we call \lq\lq intraband atomic diamagnetism''.\cite{ogata-fukuyama, ogata2,ogata3} 
$n(\mu)$ in $\chi_{\rm atomic}$ represents the total electron number with spin considered
when the chemical potential is $\mu$. 
The other contributions $\chi_{\rm LP}$, $\chi_1$ and $\chi_2$ are the primary orbital contributions.
Note that in the absense of the SOI,
$\Delta_x=\Delta_y=\Delta_{xx}=\Delta_{yy}=\Delta_{xy}=0$, so that $\chi_1$ and $\chi_2$ become simple.

\section{Numerical results\label{sec:numerical}}
\subsection{Chemical potential dependence}
\begin{figure*}
\centering\includegraphics[width=\linewidth]{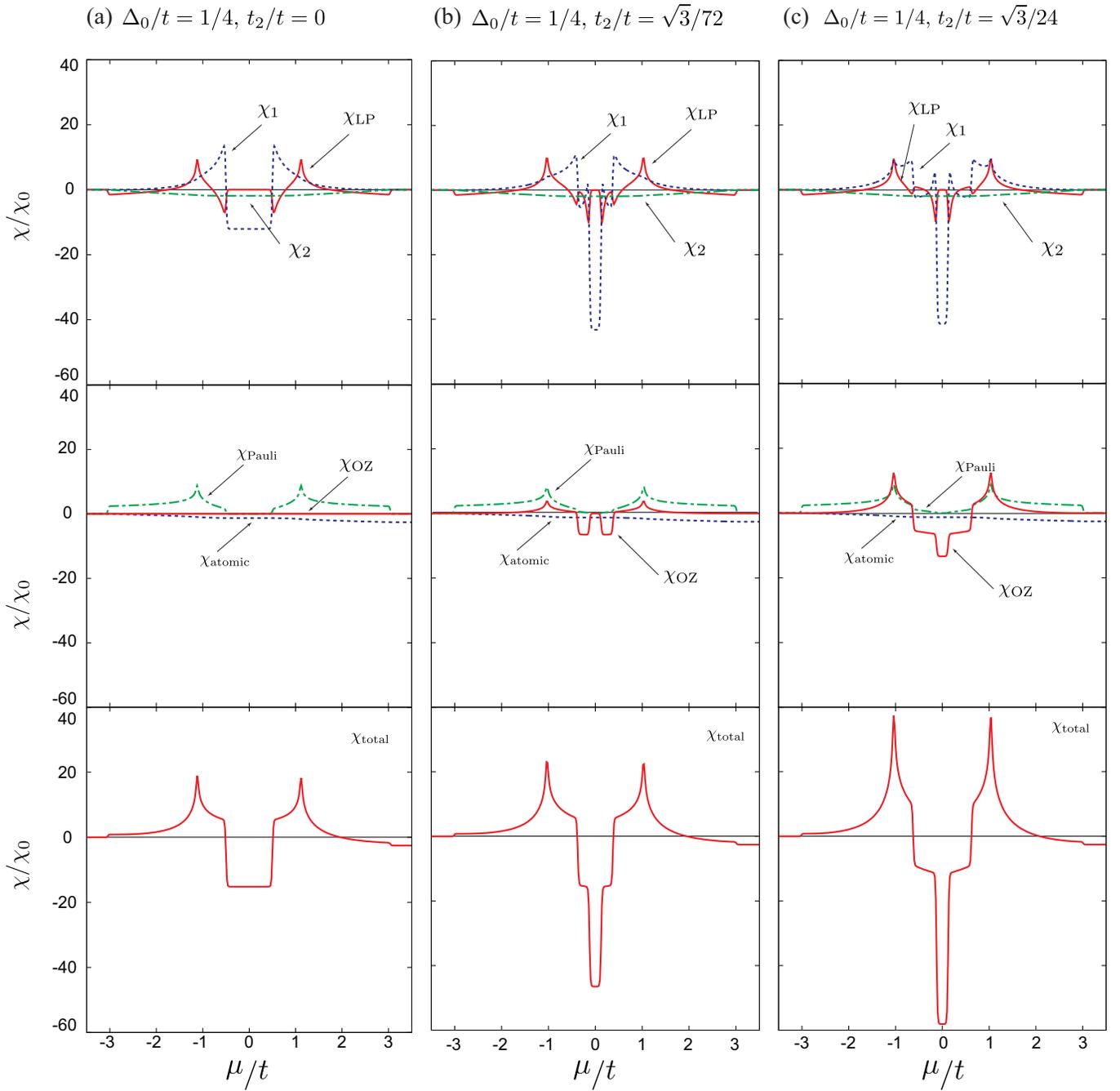}
\caption{\label{fig:chempot} 
Each contributions to the magnetic susceptibility
as a function of the chemical potential.
Top: $\chi_{\rm LP}$ (red, solid line), $\chi_1$ (blue, dashed line), and $\chi_2$ (green, dot-dashed line).
Middle: $\chi_{\rm OZ}$ (red, solid line), $\chi_{\rm atomic}$ (blue, dashed line), and $\chi_{\rm Pauli}$ (green, dot-dashed line).
Bottom: figures show the total contribution $\chi_{\rm total}$.}
\end{figure*}

Performing the numerical integration for the $\k$-summation, we obtain the magnetic susceptibility.
The parameters used are taken from the values in graphene, 
which are tabulated in Table~\ref{table:parameters} \cite{ogata3}.
Figure \ref{fig:chempot} shows each contribution to magnetic susceptibility
for some choices of parameters, $\{\Delta_0,t_2\}$:
(a) $\{\Delta_0/t, t_2/t\}=\{1/2,0\}$, (b) $\{1/4, \sqrt{3}/72\}$, and (c) $\{1/4, \sqrt{3}/24\}$ as functions of chemical potential.
In each figure, the classified contributions are shown.
The top figures show $\chi_{\rm LP}$ (red, solid line), $\chi_1$ (blue, dashed line), and $\chi_2$ (green, dot-dashed line).
The middle figures show $\chi_{\rm OZ}$ (red, solid line), $\chi_{\rm atomic}$ (blue, dashed line), and $\chi_{\rm Pauli}$ (green, dot-dashed line).
The bottom figures show the total contribution $\chi_{\rm total}$.
The values are shown in units of $\chi_0=e^2 L^2 a^2 t/(2\pi\hbar)^2$.
There are several remarks on the above results.

\begin{figure}
\centering\includegraphics[width=\linewidth]{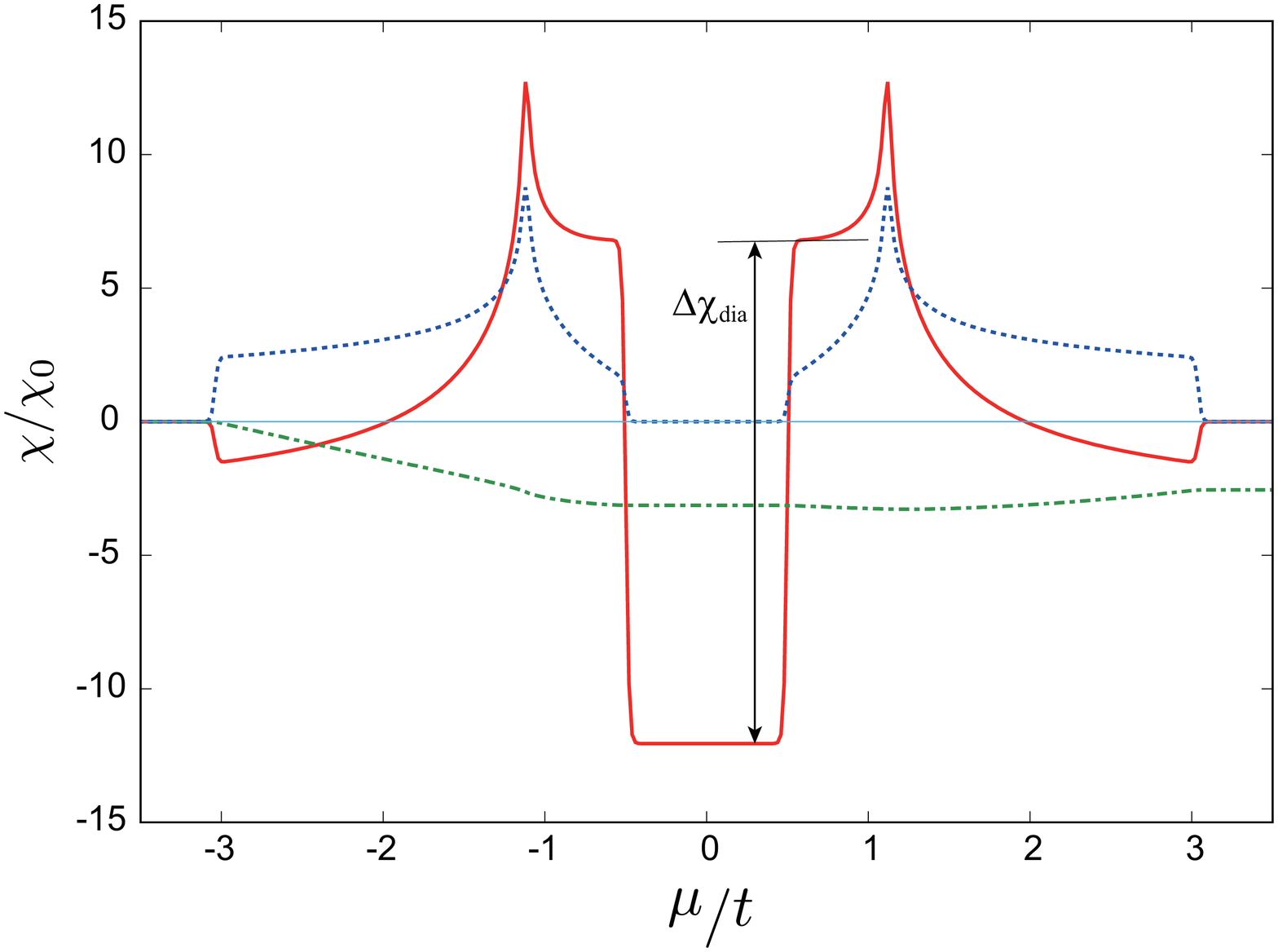}
\caption{\label{fig:raoux} Contributions to magnetic susceptibility as a function of the chemical potential
at $\Delta_0=0.5 t$ and $t_2=0$.
The solid (red), dashed (blue), and dot-dashed (green) lines correspond to 
$\chi_1+\chi_{\rm LP}$ (i.e., the result by Raoux \textit{et al.}), $\chi_{\rm Pauli}$, and $\chi_2$, respectively.
The jump at the both ends of gap in $\chi_1+\chi_{\rm LP}$ is denoted as $\Delta\chi_{\rm dia}$.
}
\end{figure}

(i) For $t_2=0$ case [Fig.~\ref{fig:chempot} (a)], $\chi_{\rm LP} + \chi_1$ coincides with the result by Raoux \textit{et al.}
\cite{raoux-piechon}, which is based on the Peierls-phase formulation.
The present result has additional contributions, $\chi_{\rm Pauli}$, $\chi_2$ and $\chi_{\rm atomic}$.
Figure~\ref{fig:raoux} shows $\chi_{\rm LP}+\chi_1$ (i.e., the result by Raoux \textit{et al.}; red, solid line), 
$\chi_{\rm Pauli}$ (blue, dahsed line), and $\chi_2+\chi_{\rm atomic}$ (green, dot-dashed line). 
$\chi_2+\chi_{\rm atomic}$ originates from the deformation of the wave functions
\cite{ogata-fukuyama,ogata2,ogata3,matsuura2016},
which is not considered in the Peierls-phase formulation.
This effect also exists for finite $t_2$ cases.
Note that $\chi_{\rm OZ}$ vanishes in this case.

(ii) The term $\chi_{\rm atomic}$ in Eq.~\eqref{eq:chi_atomic} is 
in the zeroth order 
with respect to the overlap integrals as in the square lattice and graphene cases \cite{ogata2,ogata3}.
This term is proportional to the electron number of 2$p_z$ band, $n(\mu)$.
Since this originates from the motion of an electron in an atom, 
this term does not depend on the magnitude of the overlap integral
or the amplitude of transfer integral.
This term produces asymmetric dependence on $\mu$.

(iii) At the band bottom ($\mu\simeq -\sqrt{9t^2+\Delta_0^2})$, only $\chi_{\rm LP}$ and $\chi_{\rm Pauli}$
have contributions.
The former represents the Landau-Peierls diamagnetism\cite{landau,peierls},
which is understood as the extension of Landau's diamagnetism for a periodic system;
The latter represents Pauli paramagnetism,
the magnitude of which is proportional to the density of states.
The ratio of these contribution is given by $|\chi_{\rm LP}/\chi_{\rm Pauli}|=\frac{1}{3}(m/m^*)^2$
where $m^*$ is the effective mass at the band bottom \cite{blundell}.
For $\Delta_0=0$ case, $m^*=2\hbar^2/3ta$\cite{ogata3}
and the ratio becomes $|\chi_{\rm LP}/\chi_{\rm Pauli}|=0.659$ with the parameters shown in Table \ref{table:parameters}.
Thus the total magnetic susceptibility is paramagnetic at the band bottom.

(iv) At the van Hove singularity ($\mu\simeq \pm t$), we observe sharp peaks 
in $\chi_{\rm LP}$, $\chi_{\rm Pauli}$, and $\chi_{\rm OZ}$.
This fact suggests that $\chi_{\rm OZ}$ is not just a small correction 
to the magnetic susceptibility, but one of the primary contributions.
The term $\chi_{\rm OZ}$ exists when $t_2\neq 0$ and reflects the sign of $t_2$,
and its peak at the van Hove singularity can be negative for $t_2<0$.

(v) In $\chi_{\rm total}$ near $\mu=0$ shown in Fig.~\ref{fig:chempot} (a)--(c), we find one plateau for $t_2=0$ and two plateaus for finite $t_2$.
These behaviors are explained as follows.
First, the effective Hamiltonian in the vicinities of K and K' points is given by
\begin{equation}
	H_{\rm eff}=\hbar v k_x \tau_x + \hbar v k_y \tau_y + m_0 \tau_z,
\end{equation}
where $v=3ta/2\hbar$ is velocity and $m_0=|\Delta_0+3\sqrt{3}\sigma t_2|\,(|\Delta_0-3\sqrt{3}\sigma t_2|)$ for K (K') point.
For this system, the orbital magnetic susceptibility with chemical potential $\mu$ is obtained as\cite{raoux-piechon}
\begin{align}
&\chi_{\rm 2D Dirac}(\mu,T,m_0) \nonumber \\
	&=\frac{3\pi t}{2|m_0|} (f(|m_0|)-f(-|m_0|)) \chi_0, \label{eq:diamagfintemp}\\
	&=-\frac{3\pi t}{2|m_0|} \chi_0 \theta(|m_0|-|\mu|) \quad(T=0).
	\label{eq:diamag}
\end{align}
This equation shows that $\chi_{\rm 2D Dirac}(\mu,T=0,\Delta_0)$ has a finite negative value only when 
$\mu$ is in the gap.
For finite $t_2$ cases, gaps of different sizes open at K and K' points, and thus, 
the multi-plateau structure is formed.

(vi) For the case of $t_2=\sqrt{3}/24$ (Fig~\ref{fig:chempot}(c)), 
apart from the contribution discussed in (v), we observe an extra diamagnetic contribution $\chi_{\rm OZ}$ at $\mu=0$.
This condition corresponds to the topologically nontrivial state, and $\chi_{\rm OZ}$ 
reflects the topological invariant of the model, the spin Chern number \cite{sheng}.
Note that the sign of $\chi_{\rm OZ}$ depends on $t_2$, and $\chi_{\rm OZ}$ is not always diamagnetic.
In Sec.~\ref{sec:OZ}, we discuss the relation in detail.

\begin{table}[hb]
	\caption{Parameters for graphene that are used in the numerical integration.\cite{ogata3}}
	\label{table:parameters}
	\begin{ruledtabular}
	\begin{tabular}{cd}
	\textrm{Parameters} &
		\multicolumn{1}{c}{\textrm{Value}} \\ \colrule
     $a$ &  1.42 \\ 
     $s$ & 0.237 \\  
     $t$ & 3.55 \\  
     $Z_{\rm eff}$ & 3.25 \\  
	\end{tabular}
	\end{ruledtabular}
\end{table}

\subsection{Scaling of the diamagnetic peak at $\mu=0$ \label{sec:deltafunction}}
\begin{figure}
\centering\includegraphics[width=\linewidth]{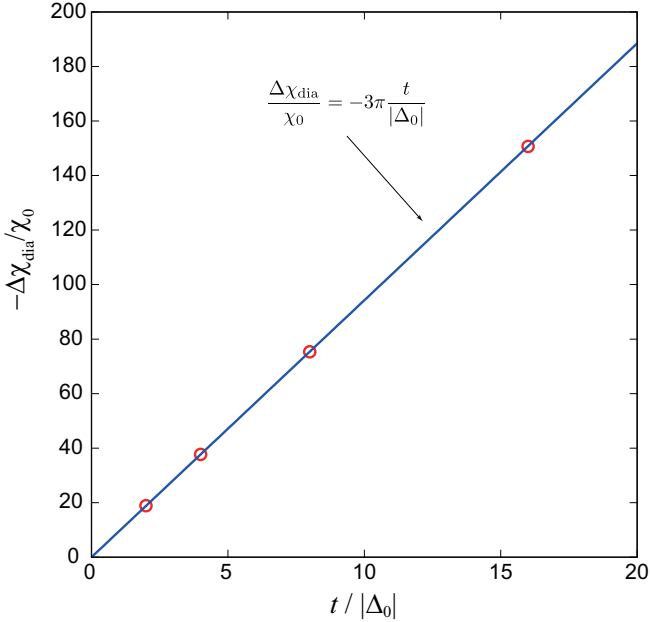}
\caption{\label{fig:ddep} Difference in magnetic susceptibility at the both ends of the gap in 
$\chi_1+\chi_{\rm LP}$, $\Delta\chi_{\rm dia}$, 
as a function of $t/\Delta_0$ at $t_2=0$ and $k_{\rm B}T/t=0.001$.
Open circles: $\Delta \chi_{\rm dia}$ obtained from our numerical result with $t/|\Delta_0|=$2.0, 4.0, 8.0, and 16.0.
Solid line: Result obtained by the continuum model Eq.~\eqref{eq:diamag}.}
\end{figure}

%
In this subsection, we discuss the jumps in the orbital magnetic susceptibility, 
i.e., $\chi_{\rm LP}+\chi_1$,
at the both ends of the gap [see Fig.~\ref{fig:raoux}]. (Note that $\chi_2$ does not contribute to jump.)
We denote the magnitude of the jump as $\Delta \chi_{\rm dia}$. 
We expect that $\Delta \chi_{\rm dia}$ should be equal to the jump $3\pi t \chi_0/|m_0|$
obtained analytically in the effective Hamiltonian with two valleys considered [see Eq.~\eqref{eq:diamag}].
The open circles in Fig.~\ref{fig:ddep} show $\Delta \chi_{\rm dia}$ obtained from our numerical result
for several values of $t/\Delta_0$ at $t_2/t=0.1$ and $k_{\rm B}T/t=0.001$.
We can see that the open circles are excellently on the line $-3\pi t \chi_0/ \Delta_0$.
(Note that $|m_0|=\Delta_0$ at $t_2=0$.)
%
%

For finite $t_2$ cases,
there are two different gaps at K and K' points, the sizes of which we denote as $\Delta^{\rm K}$ and $\Delta^{\rm K'}$, respectively.
Similarly, we find the jumps in the orbital magnetic susceptibility $\chi_{\rm LP}+\chi_1$ at the both ends of each gap,
and each jump coincides with the calculated value for the corresponding gap,
$\chi_{\rm 2D Dirac}(\mu=0,T,\Delta^{\rm K})$ or $\chi_{\rm 2D Dirac}(\mu=0,T,\Delta^{\rm K'})$.

\subsection{Relation between $\chi_{\rm OZ}$ and the topological phase \label{sec:OZ}}
We discuss the Berry curvature-related contribution $\chi_{\rm OZ}$
on the basis of the discussion previously given by the authors \cite{ozaki-ogata}.
We concentrate on the case of $T=0$ and $\mu=0$,
where the chemical potential is located in the gap.
In this case, $\chi_{\rm OZ}$ is given by
\begin{align}
	\chi_{\rm OZ}(\mu=0)=\frac{2e\mu_{\rm B}}{\hbar} \sum_{l:{\rm occupied}} 
	\sum_{\k,\sigma} \sigma \Omega_{l\sigma}^z.  \label{occ2}
\end{align}
Note that the topology of wave functions in a spin-conserved system is characterized by 
the Chern number, ${\rm Ch}_{l,\sigma}$,
where $l$ and $\sigma$ represents the band index and spin.
The Chern number is explicitly given by
\begin{equation}
	{\rm Ch}_{l,\sigma}=\frac{1}{2\pi}\int d{\bm k}\Omega_{l\sigma}
	=\frac{2\pi}{L^2}\sum_\k \Omega_{l\sigma}.
\end{equation}
and takes an integer value.
Using this relation, $\chi_{\rm OZ}$ at $\mu=0$ is given by
\begin{equation}
	\chi_{\rm OZ}(\mu=0)=-\frac{4|e|\mu_{\rm B} L^2}{h} \cdot \frac{1}{2}
	({\rm Ch}_{{\rm occ},\uparrow}-{\rm Ch}_{{\rm occ},\downarrow}).
\end{equation}
The right-hand side is proportional to the spin Chern number for the occupied band, 
a topological invariant for 2D TIs, defined by 
$({\rm Ch}_{{\rm occ},\uparrow}-{\rm Ch}_{{\rm occ},\downarrow})/2$.
This result indicates that $\chi_{\rm OZ}(\mu=0)$ is quantized in units of the universal value,
$\chi_u=4|e|\mu_{\rm B}/h=13.37\chi_0$ per area, reflecting the topological phase of materials.

\begin{figure}
\centering\includegraphics[width=\linewidth]{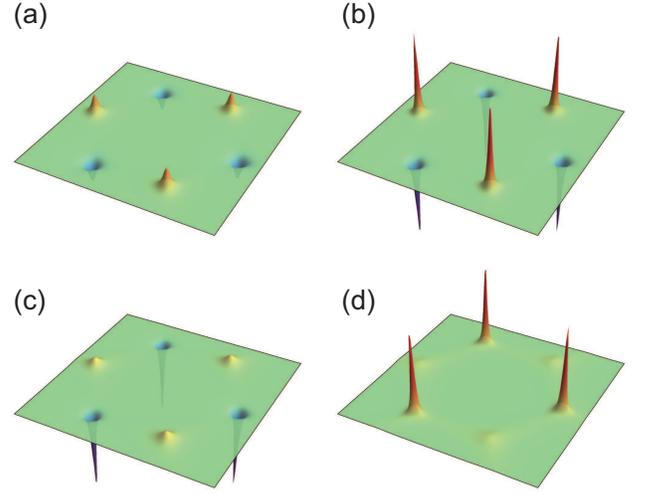}
\caption{\label{berry} 
The distribution of Berry curvature for the lower band with spin up
at each parameter: (a) $\Delta_0/t=1/4, t_2/t=0$, (b) $\Delta_0/t=1/8, t_2/t=0$,
(c) $\Delta_0/t=1/4,t_2/t=\sqrt{3}/72$, and (d) $\Delta_0/t=1/4, t_2/t=\sqrt{3}/24$.
The integral of the Berry curvature over the whole Brillouin zone is zero
when (a), (b), and (c), while the counterpart of (d) is $2\pi$.
The state is topologically non-trivial only in the case of (d).
}
\end{figure}

Figure~\ref{berry} shows the distribution of $\Omega_{l\sigma}(\k)$ for some choices of parameters.
For the case $t_2=0$ with different sizes of gaps [Fig. \ref{berry}(a,b)],
the relation of the Berry curvature $\Omega_{l\sigma}(-\k)=-\Omega_{l\sigma}(\k)$ holds
and the summation of $\Omega_{l\sigma}$ in the Brillouin zone vanishes.
For nonzero $t_2$, the relation $\Omega_{l\sigma}(-\k)=-\Omega(\k)$ does not hold
in general.
Nevertheless, as long as $|\Delta_0|>3\sqrt{3}|t_2|$ holds,
the summation of the Berry curvature in the Brillouin zone is zero and $\chi_{\rm OZ}(\mu=0)$ also vanishes [Fig.~\ref{berry}(c)].
This corresponds to the fact that the system is still topologically trivial.
On the other hand, for $|\Delta_0|<3\sqrt{3}|t_2|$, the summation becomes nonzero [Fig.~\ref{berry}(d)].
In this parameter region, the system is topologically nontrivial
and $\chi_{\rm OZ}$ has a finite contribution.
These results show that $\chi_{\rm OZ}(\mu=0)$ reflects the topological order of the Kane-Mele model.
As the ratio of $t_2$ to $\Delta_0$ changes, a jump in $\chi_{\rm OZ}(\mu=0)$ occurs
at the topological phase transition.

Let us discuss experimental detection of the jump.
For $\mu=0$, the primary contribution is $\chi_1$ as well as $\chi_{\rm OZ}$.
As shown in Sec.~\ref{sec:deltafunction}, $\chi_1$ diverges at the critical point $|\Delta_0|=3\sqrt{3}|t_2|$.
Although it seems difficult to detect the jump
due to this divergence, $\chi_{\rm OZ}$ will be experimentally observed according to the discussion below.

\begin{figure}
\centering\includegraphics[width=\linewidth]{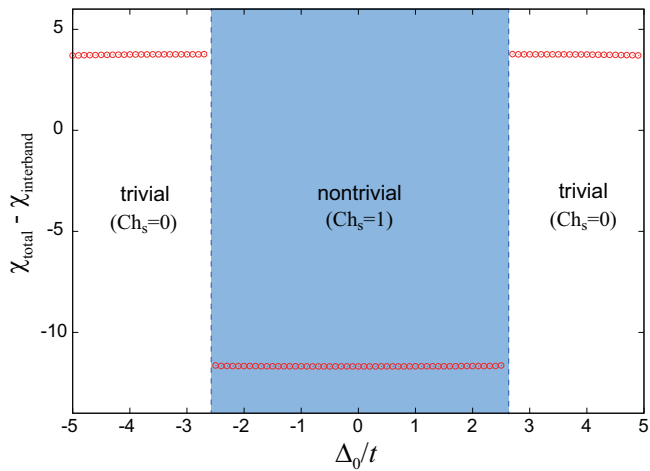}
\caption{\label{topomag} Magnetic susceptibility with the interband contribution subtracted, where we define
$\chi_{\rm interband}=\chi_{\rm 2D Dirac}(\mu=0,T,\Delta^{\rm K})+\chi_{\rm 2D Dirac}(\mu=0,T,\Delta^{\rm K'})$.
The parameters $t_2$ and $T$ are set to $t_2=0.1t$ and $k_{\rm B}T=0.001t$.
The magnitude of jump mainly comes from the universal quantized value $\chi_u=13.37\chi_0$;
however, there exist an additional discontinuous contribution originating from a \textit{purely} orbital term 
contained in $\chi_{\rm occ2}$ in Eqs.~\eqref{eq:occ2} and \eqref{eq:magmoment}.
The overall positive shift originates from the plateau shown in Fig.~\ref{fig:chempot}.}
\end{figure}

It is naturally assumed that the diverging interband contribution comes from the vicinities of 
K and K' points.
As we mentioned, we can evaluate the contribution from K and K' at $\mu=0$ as 
$\chi_{\rm 2D Dirac}(\mu=0,T,\Delta^{\rm K})+\chi_{\rm 2D Dirac}(\mu=0,T,\Delta^{\rm K'})$.
If we subtract this value from the observed total magnetic susceptibility,
we obtain the residue containing the jump in $\chi_{\rm OZ}$, i.e., the topological phase transition-related jump.
Figure~\ref{topomag} shows the residue obtained by the above subtraction as a function of $\Delta_0/t$.
In this way, we can find the evidence of a topological phase transition.
Note that the staggered on-site potential is variable by an electric field
applied perpendicularly to the 2D plane.
Therefore, the magnitude of the electric field corresponds to the horizontal axis in Fig.~\ref{topomag}.

Figure~\ref{topomag} indicates that the magnitude of the jump in the total contribution 
slightly deviates from the predicted value 13.37$\chi_0$.
This deviation originates from the Berry-curvature-related term in the \textit{purely} orbital, Berry-curvature-related
contribution contained in $\chi_{\rm occ2}$ [see Eqs.~\eqref{eq:occ2} and \eqref{eq:magmoment}],
which also changes discontinuously at topological phase transitions.
This fact does not contradict the statement that $\chi_{\rm OZ}$ is universally quantized.

Note that the effect of the Berry curvature on the orbital magnetism
was studied in some literatures 
\cite{xiao2010,thonhauser2011,sundaram,xiao2005,thonhauser2005,ceresoli,shi}.
In our formulation, the effect of the Berry curvature is included in $\chi_1$, and causes the deviation of 
the jump from quantized value.
However, this effect is purely orbital and does not affect $\chi_{\rm OZ}$.

\section{Summary \label{sec:summary}}
%
%
We calculated the orbital, spin-Zeeman, and OZ magnetic susceptibility
for the Kane-Mele model, using the formula written in terms of explicit wave functions,
which enables us to evaluate each contribution taking account of the integration 
over the whole Brillouin zone and the summation over all the bands.
The result includes additional contributions 
to the previous results \cite{ozaki-ogata}, such as core electron diamagnetism,
originating from the deformation of the wave functions by an external field.
Furthermore, the numerical calculation has revealed the following:\\
(1) The quantization of the OZ cross term is confirmed.
If we can evaluate the size of the gap with some methods, 
we will be able to detect the OZ cross term experimentally 
and observe the change in the spin Chern number directly.\\
(2) The OZ cross term can be a relatively large contribution, especially for insulating states and at the van Hove singularity, and is one of the primary contributions
to the magentic susceptibility.\\

The present study clarifies the behavior of the OZ cross term and its magnitude 
compared with the other contributions.
We expect that the OZ cross term will serve as a useful tool for experimental studies on TIs.


\begin{acknowledgments}
We thank very fruitful discussions with H.\ Matsuura, H.\ Maebashi, I.\ Tateishi, T.\ Hirosawa, N.\ Okuma, and V. K\"onye.
This work was supported by Grants-in-Aid for Scientific Research from the Japan Society for the Promotion of Science
(Grants No.~JP18H01162).
S.O. was supported by the Japan Society for the Promotion of Science through the Program for Leading Graduate Schools
(MERIT).
\end{acknowledgments}

\appendix
\begin{widetext}
\section{Hopping integrals due to the spin-orbit interaction\label{sec:spinorbit}}

In this Appendix, we obtain hopping integrals due to the spin orbit interaction using 
LCAO in Eqs.~(\ref{eq:phia}) and (\ref{eq:phib}).
%
The matrix elements of SOI in Eq.~(\ref{eq:hamfull}) 
between $\varphi_{{\rm A}{\bm k}}({\bm r})$ and  $\varphi_{{\rm B}{\bm k}}({\bm r})$
is given by
\begin{align}
	h^{\rm SO}_{{\rm AB}\sigma\sigma'}({\bm k}) &\simeq \frac{1}{N}\sum_{{\bm R}_{{\rm A}i},{\bm R}_{{\rm B}j}} 
	e^{-i\k({\bm R}_{{\rm A}i}-{\bm R}_{{\rm B}j})} \int d{\bm r} \Phi^*_\sigma ({\bm r}-{\bm R}_{{\rm A}i}) \frac{\hbar^2}{4m^2c^2} \nonumber \\
	&\times (\boldsymbol \sigma)_{\sigma\sigma'}\cdot {\bm \nabla}
	\left\{ V_A({\bm r}-{\bm R_{{\rm A}i}})+V_B({\bm r}-{\bm R_{{\rm B}j}}) \right\}
	\times \left\{ -i{\bm \nabla} \Phi_{\sigma'}({\bm r}-{\bm R}_{{\rm B}j}) \right\},
	\label{eq:a:matele-aa}
\end{align}
where we have chosen $V_{\rm A}({\bm r}-{\bm R}_{{\rm A}i})+V_{\rm B}({\bm r}-{\bm R}_{{\rm B}j})$ out of $V({\bm r})$ since
the other terms in $V({\bm r})$ should be small at ${\bm r}={\bm R}_{Ai}$ or ${\bm r}={\bm R}_{Bj}$.
We show the spin component of the wave functions explicitly.
Other matrix elements $h_{AA\sigma\sigma'}^{\rm SO}(\k)$ and $h_{BB\sigma\sigma'}^{\rm SO}(\k)$ are expressed in similar ways,
which are to be discussed later.

The dominant contribution in Eq.~(\ref{eq:a:matele-aa}) is between the nearest-neighbor-sites.
By choosing an appropriate coordinates, the ${\bm r}$ integral can be expressed 
as in Fig.~\ref{fig:spinorbit}(a). 
We can assume that $V_{\rm A}({\bm r}-{\bm R}_{{\rm A}i})$, $V_{\rm B}({\bm r}-{\bm R}_{{\rm B}j})$,
$\Phi_\sigma({\bm r}-{\bm R}_{{\rm A}i})$ and $\Phi_{\sigma'}({\bm r}-{\bm R}_{{\rm B}j})$
are even functions with respect to $y$.
Therefore, when one of the two nablas in Eq.~(\ref{eq:a:matele-aa}) is $\frac{\partial}{\partial y}$,
the ${\bm r}$ integral vanishes.
We can also assume that $V_{\rm A}({\bm r}-{\bm R}_{{\rm A}i})$ and $V_{\rm B}({\bm r}-{\bm R}_{{\rm B}j})$ 
are even functions with respect to $z$, while $\Phi_\sigma({\bm r}-{\bm R}_{{\rm A}i})$ and 
$\Phi_{\sigma'}({\bm r}-{\bm R}_{{\rm B}j})$ are odd functions.
Therefore, when one of the two nablas in Eq.~(\ref{eq:a:matele-aa}) is $\frac{\partial}{\partial z}$,
the ${\bm r}$ integral vanishes again.
As a result, $h_{{\rm AB}\sigma\sigma'}^{\rm SO}({\bm k})$ vanishes.
\begin{figure*}
\centering\includegraphics[width=\linewidth]{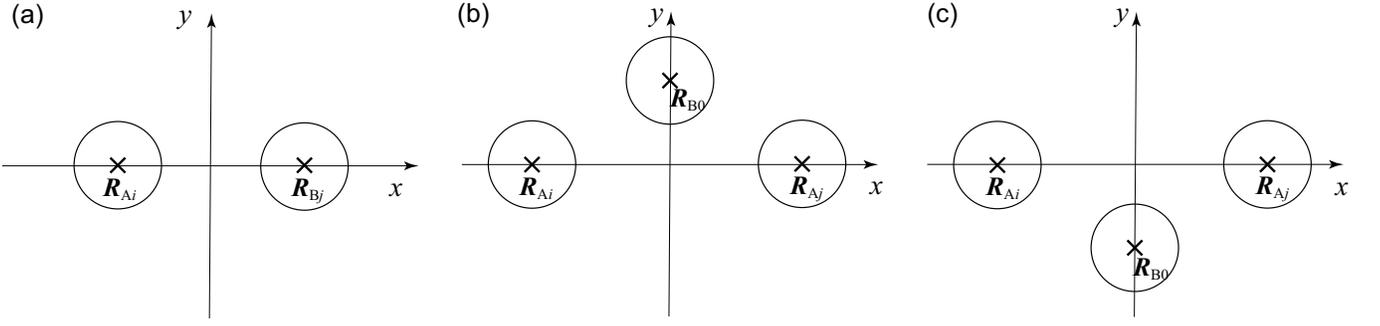}
\caption{\label{fig:spinorbit}Configurations of atoms for (a) nearest-neighbor and 
(b), (c) next-nearest-neighbor hoppings. }
\end{figure*}

Next, we discuss $h_{{\rm AA}\sigma\sigma'}^{\rm SO}({\bm k})$.
The dominant contribution is between the next-nearest-neighbor pair,
\begin{align}
	h_{{\rm AA}\sigma\sigma'}^{\rm SO}({\bm k}) &\simeq \frac{1}{N}\sum_{{\bm R}_{{\rm A}i},{\bm R}_{{\rm A}j}}
	e^{-i\k({\bm R}_{{\rm A}i}-{\bm R}_{{\rm A}j})} \int d{\bm r} 
	\Phi_{\sigma}^* ({\bm r}-{\bm R}_{{\rm A}i})
	\frac{\hbar^2}{4m^2c^2} \nonumber \\
	&\times (\boldsymbol \sigma)_{\sigma\sigma'} \cdot {\bm \nabla}
	\left\{ V_{\rm A}({\bm r}-{\bm R}_{{\rm A}i}) + V_{\rm B}({\bm r}-{\bm R}_{{\rm B}0})
	+V_{\rm A}({\bm r}-{\bm R}_{{\rm A}j}) \right\}
	\times \left\{ -i{\bm \nabla} \Phi_{\sigma'}({\bm r}-{\bm R}_{{\rm A}j}) \right\}. 
	\label{eq:a:matele-ab}
\end{align}
The appropriate coordinates give the configurations as in Fig.~\ref{fig:spinorbit}(b) and (c).
In the following, we neglect the overlap integral $s$ in $\Phi_\sigma({\bm r})$ and replace $\Phi_\sigma({\bm r})$
with $\phi_\sigma({\bm r})$.
All the potentials in Eq.~(\ref{eq:a:matele-ab}) will be even functions with respect to $z$,
while $\phi_{\sigma}^* ({\bm r}-{\bm R}_{{\rm A}i})$ and 
$\phi_{\sigma'} ({\bm r}-{\bm R}_{{\rm A}j})$ are odd functions.
Therefore, as in the case of Eq.~(\ref{eq:a:matele-aa}), when one of the two nablas is $\frac{\partial}{\partial z}$,
the ${\bm r}$ integral vanishes.
Similarly, the terms in Eq.~(\ref{eq:a:matele-ab}) with $V_{\rm A}({\bm r}-{\bm R}_{{\rm A}i})$ and $V_{\rm A}({\bm r}-{\bm R}_{{\rm A}j})$
are even functions with respect to $y$.
Therefore they vanish as in the case of Eq.~(\ref{eq:a:matele-aa}).
In contrast, the term with $V_{\rm B}({\bm r}-{\bm R}_{{\rm B}0})$ gives a nonzero value,
\begin{align}
	\frac{\hbar^2}{4m^2c^2} \int d{\bm r} \phi^*_\sigma({\bm r}-{\bm R}_{{\rm A}i}) (\sigma^z)_{\sigma\sigma'}
	\left[ \frac{\partial}{\partial x} V_{\rm B}({\bm r}-{\bm R}_{{\rm B}0}) (-i \frac{\partial}{\partial y})
	\phi_{\sigma'}({\bm r}-{\bm R}_{{\rm A}j}) \right. \nonumber \\
	\left. - \frac{\partial}{\partial y} V_{\rm B}({\bm r}-{\bm R}_{{\rm B}0}) (-i \frac{\partial}{\partial x})
	\phi_{\sigma'}({\bm r}-{\bm R}_{{\rm A}j}) \right] d{\bm r}. \label{eq:a:a3}
\end{align}
If we assume ${\bm R}_{{\rm B}0}=(0,b,0)$, 
${\bm R}_{{\rm A}j}=(a,0,0)$,
$V_{\rm B}({\bm r}-{\bm R}_{{\rm B}0})=V_{\rm B}(\rho,z)$, and 
$\phi_{\sigma'}({\bm r}-{\bm R}_{{\rm A}j})=\phi_{\sigma'}(\rho_{\rm A},z)$ 
with $\rho=\sqrt{x^2+(y-b)^2}$ and $\rho_{\rm A}=\sqrt{(x-a)^2+y^2}$ 
for Fig.~\ref{fig:spinorbit}(b), the integral in Eq.~(\ref{eq:a:a3}) becomes
\begin{equation}
-i\int d{\bm r} \phi^*_\sigma({\bm r}-{\bm R}_{{\rm A}i}) (\sigma^z)_{\sigma\sigma'}
	(ay+bx-ab)\frac{1}{\rho \rho_{\rm A}}
	\frac{\partial V_{\rm B}(\rho,z)}{\partial \rho} \frac{\partial \phi_{\sigma'}(\rho_{\rm A},z)}{\partial \rho_{\rm A}}.
	\label{eq:a:a4}
\end{equation}
Similarly, for Fig.~\ref{fig:spinorbit}(c), we obtain
\begin{align}
	&-i\int d{\bm r} \phi^*_\sigma({\bm r}-{\bm R}_{{\rm A}i}) (\sigma^z)_{\sigma\sigma'}
	(ay-bx+ab)\frac{1}{\rho' \rho_{\rm A}}
	\frac{\partial V_{\rm B}(\rho',z)}{\partial \rho'} \frac{\partial \phi_{\sigma'}(\rho_{\rm A},z)}{\partial \rho_{\rm A}},
	\label{eq:a:a5}
\end{align}
with $\rho'=\sqrt{x^2+(y+b)^2}$.
When we make the change of the integral variable $y\to -y$, we can see that Eq.~(\ref{eq:a:a5}) 
exactly equals $(-1)$ times Eq.~(\ref{eq:a:a4}).
In the same way, we can obtain $h_{\rm BB\sigma\sigma'}^{\rm SO}({\bm k})$.
These next-nearest-neighbor hoppings exactly have the same symmetry as Kane-Mele assumed,
although the absolute value and the sign is determined from the details of the functional forms in Eq.~(\ref{eq:a:a3}).



\section{Integration Formulae\label{sec:b}}

Table \ref{order-esti} shows the order estimates of several quantities with respect to the \lq\lq overlap integrals'' $s$, $t$ or $t_2$ (all denoted as $s$ in the following)
for the two cases of $\Delta_0 \gg t$ and $\Delta_0 \lesssim t$. 
Note that $E_{\k\sigma} = \sqrt{\Delta_{\k \sigma}^2 + \varepsilon_\k^2}$. 
In the following calculations, we keep the terms up to the order of $s^1$.

\begin{table}[h]
	\caption{
		Order estimations of several quantities with respect to the \lq\lq overlap integral'' 
		$s$, $t$, or $t_2$ (all denoted as $s$).
		Here $\mu$ is $x$ or $y$, and 
		$\eta_{\k\sigma}$ and $\theta_\k$ are defined in 
		Eqs.~(\ref{eq:deftheta})-(\ref{eq:defeta2}).
	}
	\begin{ruledtabular}
	\begin{tabular}{lcc}
			& $\Delta_0 \gg t$	 &  $\Delta_0 \lesssim t$ \\
			\colrule
			$E_{\k\sigma}$ & 1 & $s$ \\
			$\frac{\partial E_{\k\sigma}}{\partial k_\mu}$ & $s$ & $s$ \\
			$\eta_{\k\sigma}$ & 1 & 1 \\
			$\frac{\partial \eta_{\k\sigma}}{\partial k_\mu}$ & $s$ & 1 \\
			$\theta_\k$ & 1 & 1 \\
			$\frac{\partial \theta_\k}{ \partial k_\mu}$ & 1 & 1 \\
			$\Delta_{\k\sigma}$ & 1 & $s$ \\
			$\frac{\partial \Delta_{\k\sigma}}{\partial k_\mu}$ & $s$ & $s$ \\
	\end{tabular}
	\end{ruledtabular}
	\label{order-esti}
\end{table}

First, we show several integration formulae using $u_{\k\sigma}^\pm({\bm r})$ of Eqs.~(\ref{eq:wfplus}) and (\ref{eq:wfminus}),
which will be used in calculating $\chi$.
To simplify the expressions, we abbreviate $\varepsilon_\k$, $\theta_\k$, $E_{\k\sigma}$, 
$\Delta_{\k\sigma}$, and $\eta_{\k\sigma}$ as $\varepsilon$, $\theta$, $E$, 
$\Delta$, and $\eta$, respectively, in the following Appendices.
Furthermore, we use the abbreviations,
\begin{align}
	\epm&=\frac{\partial \ep}{\partial k_\mu}, \quad
	\theta_\mu= \frac{\partial \theta}{\partial k_\mu}, \quad
	E_\mu = \frac{\partial E}{\partial k_\mu}, \quad
	\dksm=\frac{\partial \dks}{\partial k_\mu}, \quad
	\eta_\mu = \frac{\partial \eta}{\partial k_\mu},\\
	\epmn&=\frac{\partial^2 \ep}{\partial k_\mu \partial k_\nu}, \quad
	\theta_{\mu\nu}=\frac{\partial^2 \theta}{\partial k_\mu \partial k_\nu}, \quad
	E_{\mu\nu}=\frac{\partial^2 E}{\partial k_\mu \partial k_\nu}, \quad
	\dksmn=\frac{\partial^2 \dks}{\partial k_\mu \partial k_\nu}, \quad
	\eta_{\mu\nu}=\frac{\partial^2 \eta}{\partial k_\mu \partial k_\nu}, 
\end{align}
for $\mu,\nu=x\,{\rm or}\, y$. Then we obtain
\begin{align}
&{\rm (F1)}	\int \upms \frac{\partial \upm}{\partial k_\mu} \dr = \pm i\frac{\dks}{2\ek} \theta_\mu 
	+O(s^2),\nonumber \\
&{\rm (F2)}	\int \upms \frac{\partial \ump}{\partial k_\mu} \dr = 
i\frac{\ep}{2\ek}\theta_\mu \pm \eta_\mu +O(s^2), \nonumber \\
&{\rm (F3)}	\int \frac{\partial \upms}{\partial k_\mu} \frac{\partial \upm}{\partial k_\nu} d{\bm r}=
	\langle x_\mu x_\nu \rangle + \frac{1}{4}\theta_\mu \theta_\nu + \eta_\mu \eta_\nu 
	\pm i \frac{\ep}{2\ek}(\them\eta_\nu - \then\eta_\mu)
	\mp \frac{\ep}{\ek}{\rm Re}X_{\mu\nu} +O(s^2),\nonumber \\
&{\rm (F4)}	\int \frac{\partial \upms}{\partial k_\mu} \frac{\partial \ump}{\partial k_\nu} d{\bm r}= 
	-i \frac{\dks}{2 \ek}(\them \eta_\nu - \then \eta_\mu) + Y^\pm_{\mu\nu}
	+O(s^2),\nonumber \\
&{\rm (F5)}	\int \upms \Hmu \upm \dr = \pm E_\mu, \nonumber \\
&{\rm (F6)}	\int \upms \Hmu \ump \dr = \mp 2E \int u_{\k\sigma}^{\pm *}  
\frac{\partial u^{\mp}_{\k\sigma}}{\partial k_\mu}d{\bm r} 
	=\mp i\ep \them - 2E\eta_\mu +O(s^2), \nonumber \\
&{\rm (F7)}	\int \upms \Hmu \frac{\partial \upm}{\partial k_\nu} \dr = -\frac{\hbar^2}{2m} \delta_{\mu\nu}
	\pm \frac{1}{2} E_{\mu\nu} + i \frac{\dks}{2\ek} E_\mu \then
	-i\varepsilon (\eta_\mu \theta_\nu - \eta_\nu \theta_\mu) +O(s^2), \nonumber \\
&{\rm (F8)}	\int \umps \Hmu \frac{\partial \upm}{\partial k_\nu} \dr =
	\pm \frac{i}{2} \epn\them -\eta_\mu E_\nu - E\eta_{\mu\nu} 
	\mp E Y^\mp_{\mu\nu} \pm \frac{i}{2}\ep\theta_{\mu\nu} +O(s^2), 
\end{align}
with
\begin{align}
	X_{\mu\nu} &= \sum_{\R} e^{-i\theta_\k} e^{-i\k\cdot \R} \langle x_\mu x_\nu \rangle_{\bm R},
	\nonumber \\
	Y_{\mu\nu}^\pm &= \frac{\dks}{\ek} {\rm Re} X_{\mu\nu} \pm i {\rm Im} X_{\mu\nu}.
\end{align}
The meaning of the ${\bm R}$ summation and $\langle \cdots \rangle_{\bm R}$ are shown below.
$X_{\mu\nu}$ and $Y^\pm_{\mu\nu}$ are in the order of $s$. 

\subsection{Derivation of (F1)-(F4)}

The $k_\mu$ derivative of $u^+_{\k\sigma}$ becomes 
\begin{align}
	\frac{\partial u^+_{\k\sigma}}{\partial k_\mu} = 
	\frac{i}{2}\theta_\mu \cos 2\eta u^+_{\k\sigma} + \left (\frac{i}{2}\them \sin 2\eta - \eta_\mu \right ) u^-_{\k\sigma} 
	+ e^{\frac{i}{2}\theta} \cos \eta \frac{\partial \varphi_{{\rm A}\k}}{\partial k_\mu}
	- e^{-\frac{i}{2}\theta} \sin \eta \frac{\partial \varphi_{{\rm B}\k}}{\partial k_\mu}.
\end{align}
To obtain (F1), we must calculate the integral of product of Bloch wavefunction $\upm$ and 
$k$-derivative of $\varphi_{{\rm A/B}\k}(\r)$. They become
\begin{align}
	\int u_{\k\sigma}^{+*} \frac{\partial \varphi_{{\rm A}\k}}{\partial k_\mu}\dr &=
	-e^{-\frac{i}{2}\theta} \cos \eta \frac{i}{N} \sum_{\RAi,\RAj}e^{-i\k(\RAi-\RAj)} 
	\int (x-R_{Ajx})\Phi^* (\r-\RAi) \Phi(\r-\R_{Aj}) \dr \nonumber \\
	&+e^{\frac{i}{2}\theta} \sin \eta \frac{i}{N} \sum_{\RAj,\RBi}e^{i\k(\RAj-\RBi)}
	\int (x-R_{Ajx})\Phi^* (\r-\RBi) \Phi(\r-\R_{Aj}) \dr.
\end{align}
Hereafter, we only consider the on-site and adjacent-site contributions. Then we obtain
\begin{align}
	\int u_{\k\sigma}^{+*} \frac{\partial \varphi_{{\rm A}\k}}{\partial k_\mu}\dr &=
	-e^{-\frac{i}{2}\theta} \cos \eta \frac{i}{N} \sum_{\RAi}
	\int (x-R_{Aix})\Phi^* (\r-\RAi) \Phi(\r-\R_{Ai}) \dr \nonumber \\
	&+e^{\frac{i}{2}\theta} \sin \eta \frac{i}{N} 
	\sum_{\substack{\RAj, \RBi \\ {\bm R}_{{\rm A}j}-{\bm R}_{{\rm B}i}=n.n.}}
	e^{i\k(\RAj-\RBi)}	\int (x-R_{Ajx})\Phi^* (\r-\RBi) \Phi(\r-\R_{Aj}) \dr \nonumber \\
	&=-i e^{-\frac{i}{2}\theta} \cos \eta \langle x \rangle
	+ i\sin \eta \sum_\R e^{i\k\R} \langle x \rangle _{(-{\bm R})},
\end{align}
where ${\bm R}$ runs over the three vectors from a B site to its adjacent A sites.
The expectation values $\langle O \rangle$ and $\langle O \rangle_\R$ 
are defined as \cite{ogata2,ogata3}
\begin{align}
	\langle O \rangle = \int \Phi^* (\r) \hat{O} \Phi(\r) \dr,
\end{align}
and
\begin{align}
	\langle O \rangle _\R = \int \dr \Phi^*(\r-\R) \hat{O} \Phi(\r) \dr.
\end{align}
We can see that $\langle x \rangle$ and $ \langle x \rangle_\R$ are in the order of $s^2$. 
Therefore, we obtain
\begin{align}
	\int u_{\k\sigma}^{+*} \frac{\partial \varphi_{{\rm A}\k}}{\partial k_\mu} \dr = O(s^2). \label {udpa}
\end{align}
For $\partial \varphi_{{\rm B}\k}/\partial k_\mu$, we have the similar relation
\begin{align}
	\int u^{+*}_{\k\sigma} \frac{\partial \phib}{\partial k_\mu} \dr = O(s^2). \label {udpb}
\end{align}

Using Eqs. (\ref{udpa}) and (\ref{udpb}), we have
\begin{align}
  \int u^{+*}_{\k\sigma} \frac{\partial u^+_{\k\sigma}}{\partial k_\mu} \dr = \frac{i}{2} \theta_\mu \cos 2\eta + O(s^2),
 \end{align}
and
\begin{align}
  \int u_{\k\sigma}^{+*} \frac{\partial u^-_{\k\sigma}}{\partial k_\mu} \dr = \frac{i}{2} \them \sin 2\eta + \eta_\mu + O(s^2).
\end{align}
If we substitute $\eta \rightarrow \eta - \pi/2$, $u^+_{\k\sigma}$ and $u^-_{\k\sigma}$ 
become $u^-_{\k\sigma}$ and $-u^+_{\k\sigma}$, respectively.
With the relations
\begin{align}
  \sin 2\eta=\frac {\varepsilon}{E} ,  \quad \cos 2\eta=\frac{\Delta}{E}, 
\end{align}
we obtain (F1) and (F2).

%
Similarly to the derivation of (F1) and (F2), we obtain
\begin{align}
  \int \frac{\partial \phia^*}{\partial k_\mu} \frac{\partial \phia}{\partial k_\nu} \dr
  =\int \frac{\partial \phib^*}{\partial k_\mu} \frac{\partial \phib}{\partial k_\nu} \dr
  = \langle x_\mu x_\nu \rangle + O(s^2),
\end{align}
and
\begin{align}
  \int \frac{\partial \phia^*}{\partial k_\mu} \frac{\partial \phib}{\partial k_\nu} \dr
  = \sum_\R e^{-i\k\cdot \R} \langle x_\mu x_\nu \rangle_\R + O(s^2).
\end{align}
Using these relations and substitution of $\eta \rightarrow \eta - \pi/2$,
we obtain (F3) and (F4).
In the different sign cases (e.g., $\int \frac{\partial u^{+*}_{\k\sigma}}{\partial k_\mu}\frac{\partial u^-_{\k\sigma}}{\partial k_\nu}\dr$),
we can calculate the integral in almost the same way.

\subsection{Derivation of (F5) and (F6)}
We start from the Schr$\ddot {\rm o}$dinger equation,
\begin{align}
  H_\k \upm=\pm E \upm.
\end{align}
Differentiating the both sides of this equation by $k_\mu$, we obtain
\begin {align}
  \left(\Hmu \mp \frac{\partial E}{\partial k_\mu} \right)\upm
  = (\pm E - H_\k) \frac{\partial \upm}{\partial k_\mu} \label{dschroe}.
\end{align}
When we multiply $\upms$ and integrate the product, we obtain (F5).
Similarly, multiplying $\umps$ and using (F2), we obtain (F6).

\subsection{Derivation of (F7)}
To obtain (F7), we first calculate
\begin{align}
  \int \upms \left(\Hmu \frac{\partial \upm}{\partial k_\nu} 
  + \Hnu \frac{\partial \upm}{\partial k_\mu} \right) \dr \label{2diff-plus}
\end{align}
and
\begin{align}
  \int \upms \left(\Hmu \frac{\partial \upm}{\partial k_\nu}
   - \Hnu \frac{\partial \upm}{\partial k_\mu} \right) \dr. \label{2diff-minus}
\end{align}
Differentiating the both sides of Eq.~(\ref{dschroe}) by $k_\nu$, we obtain
\begin{align}
  \left ( \frac{\hbar^2}{m} \delta_{\mu\nu} \mp \frac{\partial^2 E}{\partial k_\mu \partial k_\nu} \right) \upm
  +\left( \Hmu \mp \frac{\partial E}{\partial k_\mu} \right) \frac{\partial \upm}{\partial k_\nu}
  +\left( \Hnu \mp \frac{\partial E}{\partial k_\nu} \right) \frac{\partial \upm}{\partial k_\mu}
  +(H_\k \mp E) \frac{\partial ^2 \upm}{\partial k_\mu \partial k_\nu}=0. \label{2nd-diff}
\end{align}
Here we have used the relation
\begin{align}
  \frac{\partial^2 H_\k}{\partial k_\mu \partial k_\nu}=\frac{\hbar^2}{m} \delta_{\mu\nu}.
\end{align}
Then, multiplying $\upms$ and integrating, we obtain
\begin{align}
  \int \upms \left(\Hmu \frac{\partial \upm}{\partial k_\nu} 
  + \Hnu \frac{\partial \upm}{\partial k_\mu} \right) \dr 
  =-\frac{\hbar^2}{m}\delta_{\mu\nu} \pm E_{\mu\nu} 
  + i\frac{\Delta}{2E} (E_\mu \then + E_\nu \them) +O(s^2).
  \label{2diff-plus-answer}
\end{align}
Here we have used the formula (F1).

Next, we calculate Eq.~(\ref{2diff-minus}). 
By using the explicit forms of $\Hy$ and $\frac{\partial \upm}{\partial k_x}$ 
and using the fact that $\phi(\r)$ ($p_z$-orbital) is an eigenstate of 
angular momentum, $L_z \phi(\r)=0$, we can write
\begin{align}
  \Hy \frac{\partial \upm}{\partial k_x} - \Hx \frac{\partial \upm}{\partial k_y}
  =&\Hy \left [ \pm \left (i\frac{\thex}{2}\frac{\dks}{\ek}+\frac{s\epx}{2t} \frac{\ep}{\ek} \right ) \upm
   + \left (i\frac{\thex}{2} \frac{\ep}{\ek} \mp \eta_x \pm \frac{is\ep}{2t}\thex 
	 - \frac{s \epx}{2t} \frac{\dks}{\ek} \right )\ump
	 \right ] \nonumber \\
   &- (x \leftrightarrow y)+O(s^2).
	\label{eq:dd-dd}
\end{align}
Then we multiply $\upms$ and integrate the product. After some algebra, we obtain
\begin{align}
  \int \upms \left(\Hmu \frac{\partial \upm}{\partial k_\nu} 
  - \Hnu \frac{\partial \upm}{\partial k_\mu} \right) \dr
  =i \frac{\Delta}{2E} (E_\mu \then - E_\nu \them)
  - 2i\varepsilon (\eta_\mu \then - \eta_\nu\them)
	+O(s^2). \label{2diff-minus-answer}
\end{align}
Here, we have used the formulae (F5), and (F6).
Combining Eqs.~(\ref{2diff-plus-answer}) and (\ref{2diff-minus-answer}), we obtain (F7).

\subsection{Derivation of (F8)}
Similarly to the derivation of (F7), we calculate
\begin{align}
  \int \umps \left(\Hmu \frac{\partial \upm}{\partial k_\mu} 
  + \Hnu \frac{\partial \upm}{\partial k_\nu} \right) \dr,
\end{align}
and
\begin{align}
  \int \umps \left(\Hmu \frac{\partial \upm}{\partial k_\mu}
   - \Hnu \frac{\partial \upm}{\partial k_\nu} \right) \dr.
\end{align}

First, we multiply $\umps$ to Eq.(\ref{2nd-diff}) and integrate the product. Then we obtain
\begin{align}
  \int \umps \left( \Hmu \frac{\partial \upm}{\partial k_\nu} + 
  \Hnu \frac{\partial \upm}{\partial k_\mu} \right) \dr =
  \pm E_\mu \int \umps \frac{\partial \upm }{\partial k_\nu}\dr
  \pm E_\nu \int \umps \frac{\partial \upm }{\partial k_\mu} \dr
  \pm 2\ek \int \umps \frac{\partial^2 \upm}{\partial k_\mu \partial k_\nu} \dr
	\label{eq:intminusplus}
\end{align}
To calculate the last term, by differentiating (F2) by $k_\nu$, we find a relation,
\begin{align}
  \int \upms \frac{\partial^2 \ump}{\partial k_\mu \partial k_\nu} \dr
 & =-\int \frac{\partial \upms}{\partial k_\mu} \frac{\partial \ump}{\partial k_\nu} \dr
  + \frac{\partial}{\partial k_\mu} \int \upms \frac{\partial \ump}{\partial k_\nu} \dr
  \nonumber \\
  &=i \frac{\Delta}{2\ek} (\them\eta_\nu + \then\eta_\mu) + i\frac{\ep}{2\ek}\theta_{\mu\nu}
  \pm \eta_{\mu\nu} - Y_{\mu\nu}^\pm +O(s^2),
\end{align}
where we have used a relation Eq.~(\ref{eq:SpRel0}) or Eq.~(\ref{eq:SpRel1}).
Substitution of this relation to Eq.~(\ref{eq:intminusplus}) leads to
\begin{align}
  &\int \umps \left( \Hmu \frac{\partial \upm}{\partial k_\nu} 
  + \Hnu \frac{\partial \upm}{\partial k_\mu} \right) \dr \nonumber \\
  &= \pm \frac{i}{2} (\epm \then + \epn \them) -( E_\mu\eta_\nu + E_\nu \eta_\mu) 
  \pm i\ep \theta_{\mu\nu} \mp 2E Y^\mp_{\mu\nu}
  - 2E \eta_{\mu\nu} +O(s^2).
  \label{2diff-plus-answer2}
\end{align}

On the other hand, using Eq.~(\ref{eq:dd-dd}), we obtain
\begin{align}
  &\int \umps \left( \Hmu \frac{\partial \upm}{\partial k_\nu} 
  - \Hnu \frac{\partial \upm}{\partial k_\mu} \right) \dr \nonumber \\
  &= \pm \left (i\frac{\Delta \then}{2E} + \frac{s \varepsilon \epn}{2tE} \right )
  \int \umps \Hmu \upm \dr - (\mu \leftrightarrow \nu) \nonumber \\
  &\quad + \left (i\frac{\varepsilon \then}{2E} \mp \eta_\nu \pm \frac{is\varepsilon}{2t}\then
	 - \frac{s\Delta \epn}{2tE} \right )
  \int \umps \Hmu \ump \dr - (\mu \leftrightarrow \nu) \nonumber \\
  &=\mp \frac{i}{2} (\epm\then - \epn\them)
	-\frac{1}{2\ek}(\epm\Delta_\nu - \epn\Delta_\mu) + O(s^2),
	  \label{2diff-minus-answer2}
\end{align}
where we have used (F5), and (F6).
Combining Eqs.~(\ref{2diff-plus-answer2}) and (\ref{2diff-minus-answer2}), we obtain (F8).
\end{widetext}

\section{Several relations between momentum derivatives}

We find various relations between $E_\mu, \eta_\mu, \varepsilon_\mu, \Delta_\mu$, 
and $\theta_\mu$, 
which are used in various occasions. 
Using $\sin 2\eta=\varepsilon/E$ and $\cos 2\eta=\Delta/E$, we can see
\begin{equation}
\frac{\partial}{\partial k_\mu} \left( \frac{\varepsilon}{E} \right)
= \frac{\partial}{\partial k_\mu} \sin 2 \eta
= 2 \eta_\mu \cos 2\eta 
= \frac{2\Delta}{E}\eta_\mu.
\label{eq:SpRel0}
\end{equation}
By writing explicitly the left-hand side, we obtain
\begin{equation}
\Delta \eta_\mu + \frac{\varepsilon}{2E} E_\mu = \frac{\varepsilon_\mu}{2}.
\label{eq:SpRel1}
\end{equation}
Similarly, from the derivative of $\Delta/E=\cos 2\eta$, we have
\begin{equation}
-\varepsilon \eta_\mu + \frac{\Delta}{2E} E_\mu = \frac{\Delta_\mu}{2},
\label{eq:SpRel2}
\end{equation}
which was used in the text Eq.~(\ref{eq:SpRel1text}).
Furthermore, by making the $k_\nu$-derivative of Eqs.~(\ref{eq:SpRel1}) and (\ref{eq:SpRel2}),
we obtain
\begin{equation}
E\eta_{\mu\nu} + E_\mu \eta_\nu + E_\nu \eta_\mu 
= \frac{1}{2E}(\Delta \varepsilon_{\mu\nu} - \varepsilon \Delta_{\mu\nu}).
\label{eq:SpRel3}
\end{equation}
and
\begin{equation}
E_{\mu\nu} - 4E \eta_\mu \eta_\nu 
= \frac{1}{E}(\varepsilon \varepsilon_{\mu\nu} + \Delta \Delta_{\mu\nu}).
\label{eq:SpRel4}
\end{equation}

In the previous paper, we obtained the following relationships:\cite{ogata3}
\begin{align}
&\varepsilon (\theta_x^2+\theta_y^2) - \varepsilon_{xx} - \varepsilon_{yy}
= a^2 \varepsilon, \nonumber\\
&\varepsilon (\theta_{xx}+\theta_{yy}) +2\varepsilon_{x}\theta_x + 2\varepsilon_{y}\theta_y
=0,
\label{eq:AppCnew1}
\end{align}
with $a$ being the length between the nearest-neighbor carbons.
These special relations hold since $\varepsilon$ and $\theta$ are closely related 
to each other through $\gamma_{\bm k}$. 
Here we find additional relations.

Let us consider 
\begin{equation}
\sum_{{\bm R}} (R_{x}^2-R_{y}^2) e^{-i{\bm k}\cdot {\bm R}},
\end{equation}
where $\bm R$ runs over the three vectors from a B site to its adjacent A sites. 
By using the explicit three vectors in Fig.~1, we can see that it is equal to
\begin{equation}
-ia \frac{\partial}{\partial k_y} \sum_{{\bm R}} e^{-i{\bm k}\cdot {\bm R}}.
\end{equation}
Then, from the definitions of $\gamma_{\bm k}$ and $\theta$, we can obtain the relationship:
\begin{equation}
\left( -\frac{\partial^2}{\partial k_x^2} + \frac{\partial^2}{\partial k_y^2} \right) 
| \gamma_{\bm k}| e^{i\theta} = -ia \frac{\partial}{\partial k_y}| \gamma_{\bm k}| e^{i\theta}.
\end{equation}
By taking the real and imaginary part of both sides, we obtain
\begin{align}
&\varepsilon (\theta_x^2-\theta_y^2) - \varepsilon_{xx} + \varepsilon_{yy}
= a \varepsilon \theta_y, \nonumber\\
&\varepsilon (\theta_{xx}-\theta_{yy}) +2\varepsilon_{x}\theta_x - 2\varepsilon_{y}\theta_y
=a\varepsilon_y.
\label{eq:AppCnew2}
\end{align}
From Eqs.~(\ref{eq:AppCnew1}) and (\ref{eq:AppCnew2}), we can see
\begin{align}
\varepsilon_{xx} &=\varepsilon \theta_x^2- \frac{a^2}{2} \varepsilon 
-\frac{a}{2} \varepsilon \theta_y, \nonumber\\
\varepsilon_{yy} &=\varepsilon \theta_y^2- \frac{a^2}{2} \varepsilon 
+\frac{a}{2} \varepsilon \theta_y, \nonumber\\
\varepsilon \theta_{xx} &= -2\varepsilon_x \theta_x + \frac{a}{2} \varepsilon_y, \nonumber \\
\varepsilon \theta_{yy} &= -2\varepsilon_y \theta_y - \frac{a}{2} \varepsilon_y,
\label{eq:AppCnew3}
\end{align}
Similarly by using 
\begin{equation}
\sum_{{\bm R}} R_{x}R_{y} e^{-i{\bm k}\cdot {\bm R}}
=-\frac{\partial^2}{\partial k_x\partial k_y} 
| \gamma_{\bm k}| e^{i\theta} = -i \frac{a}{2} \frac{\partial}{\partial k_x} 
| \gamma_{\bm k}| e^{i\theta},
\end{equation}
we obtain
\begin{align}
\varepsilon_{xy} &=\varepsilon \theta_x \theta_y - \frac{a}{2} \varepsilon \theta_x, 
\nonumber\\
\varepsilon \theta_{xy} &= -\varepsilon_x \theta_y - \varepsilon_y \theta_x 
+ \frac{a}{2}\varepsilon_x.
\label{eq:AppCnew4}
\end{align}

\begin{widetext}
\section{Each contribution of $\chi$\label{sec:c}}

In the present case, the Landau-Peierls contribution simply becomes Eq.~(\ref{eq:chi_LP}).
Next, the $l'$-summation in $\chi_{\rm inter}$ is carried out in Appendix \ref{sec:chiinter}, and 
the result is given by
\begin{align}
  &\chi_{\rm inter} = \frac{e^2}{\hbar^2} \sum f(\pm\ek) \nonumber \\
  & \times \biggl[  \left (\frac{\hbar^2}{16m} \pm \frac{\ep^2}{4\ek} \rsq
  \pm \frac{\hbar^2}{8m} \frac{s\ep^2}{\ek t} \right ) \thex^2 
  + \frac{\hbar^2}{4m} \left(\eta_x^2 \pm \frac{s\Delta}{Et} \eta_x \varepsilon_x \right)
  \pm E \rsq \eta_x^2 \nonumber \\  
  & \pm \frac{\eta_x^2}{4\ek}(\varepsilon\epyy +\Delta\Delta_{yy}) 
    \mp \frac{\eta_x \eta_y}{4\ek} (\varepsilon \epxy + \Delta\Delta_{xy}) 
   \pm \frac{\thex^2}{16\ek}(\ep\epyy-\Delta\Delta_{yy}) 
   \mp \frac{\thex \they}{16\ek} (\ep\epxy-\Delta\Delta_{xy})\nonumber \\
  &\pm \frac{\thex^2}{4\ek}(\Delta \epy \eta_y -E^2 \eta_y^2) 
  \mp \frac{\thex\they}{4\ek} (\Delta\epx\eta_y-E^2 \eta_x \eta_y) 
  \mp \frac{\ep\Delta}{4\ek}(\eta_x \thex\theyy - \eta_x \they\thexy) 
  \biggr] + (x\leftrightarrow y) +O(s^2). 
  \label{eq:chiInterFinal}
\end{align}

For $\chi_{\rm FS}$, we multiply the Hermitian conjugate of Eq.~(\ref{eq:dd-dd}) by 
$\frac{\partial u_{\k\sigma}^{\pm}}{\partial k_y}$ and integrate the product.
Then, with the help of (F7) and (F8), we obtain
\begin{align}
	\chi_{\rm FS}&=\frac{e^2}{2\hbar^2} \sum_{\pm\k\sigma} f'(\pm E) 
	\biggl[ \frac{\hbar^2}{2m} \frac{s\varepsilon}{2tE} \varepsilon_x E_x
	+ \frac{\varepsilon^2 E_x}{4E} (\they\thexy - \thex\theyy) 
	+\frac{\Delta E_x}{4E}(\Delta_x \they^2 - \Delta_y\thex\they) \nonumber \\
	&+ \frac{E_x}{2E} \left\{ \eta_y (\Delta \varepsilon_{xy}-\varepsilon \Delta_{xy}) - \eta_x (\Delta\varepsilon_{yy}-\varepsilon \Delta_{yy})\right\} 
	+ E_x^2 \left( \langle y^2 \rangle+\frac{1}{4}\theta_y^2 \right) -\frac{1}{4}E_xE_y \theta_x \theta_y
	+(x\leftrightarrow y)  \biggr] \nonumber \\
	&+\frac{e^2}{2\hbar^2} \sum_{\pm\k\sigma} f'(\pm E) \frac{\hbar^2}{m} \frac{\Delta}{2E}
	\sigma \left( E_x \theta_y - E_y \theta_x \right) + O(s^2),
\label{eq:c2}
\end{align}
where we have used the relation in Eqs.~(\ref{eq:SpRel2}) and (\ref{eq:SpRel3}).
The last term in Eq.~(\ref{eq:c2}) comes from the contribution of the Zeeman term.
We find that it is convenient to make partial integrations in the last three terms, which 
will be included in $\chi_1, \chi_2$, and $\chi_{\rm OZ}$ in the main text. 
With the partial integrations, we obtain
\begin{align}
	\chi_{\rm FS}&=\frac{e^2}{2\hbar^2} \sum_{\pm\k\sigma} f'(\pm E) 
	\biggl[ \frac{\hbar^2}{2m} \frac{s\varepsilon}{2tE} \varepsilon_x E_x
	+ \frac{\varepsilon^2 E_x}{4E} (\they\thexy - \thex\theyy) 
	+\frac{\Delta E_x}{4E}(\Delta_x \they^2 - \Delta_y\thex\they) \nonumber \\
	&+ \frac{E_x}{2E} \left\{ \eta_y (\Delta \varepsilon_{xy}-\varepsilon \Delta_{xy}) - \eta_x (\Delta\varepsilon_{yy}-\varepsilon \Delta_{yy})\right\} 
	+(x\leftrightarrow y)  \biggr] \nonumber \\
	&-\frac{e^2}{2\hbar^2} \sum_{\pm\k\sigma} f(\pm E) 
	\biggl[ \pm E_{xx} \left( \langle y^2 \rangle+\frac{1}{4}\theta_y^2 \right) 
	\mp \frac{1}{4}E_{xy} \theta_x \theta_y
	\mp \frac{1}{4} \left( E_x \theta_x \theta_{yy}-E_x \theta_y \theta_{xy} \right)
	+(x\leftrightarrow y)  \biggr] \nonumber \\
	&-\frac{e^2}{2\hbar^2} \sum_{\pm\k\sigma} f(\pm E) \frac{\hbar^2}{m} \sigma
	\left\{ \left( \frac{\Delta_x}{2E}-\frac{\Delta E_x}{2E^2} \right) \theta_y 
	- \left( \frac{\Delta_y}{2E}-\frac{\Delta E_y}{2E^2} \right) \theta_x \right\}  + O(s^2),
\label{eq:c22}
\end{align}

$\chi_{\rm FS-P}$ is directly obtained by substituting $M_{\pm\pm \sigma}$ as
\begin{equation}
   \chi_{\rm FS-P} = -\sum_{\pm,\k,\sigma} f'(\pm E) 
\left[ \frac{e^2}{4\hbar^2} (\Delta_x \theta_y -\Delta_y \theta_x)^2
+\frac{e^2}{2m} \sigma  (\Delta_x \theta_y -\Delta_y \theta_x)
+\frac{e^2\hbar^2}{4m^2} \sigma^2 \right] + O(s^2).
\end{equation}
Using the integration formulae (F3) in Appendix~\ref{sec:b}, we obtain $\chi_{\rm occ1}$ as
\begin{align}
%
  \chi_{\rm occ1} = -\frac{e^2}{4\hbar^2} \sum_{\pm\k\sigma} f(\pm E)
  &\biggl[ \pm E_{xy} \left(\frac{1}{4} \theta_x \theta_y + \eta_x \eta_y \right) 
  +\left( \frac{\hbar^2}{m} \mp E_{xx} \right) 
  \left(\langle y^2 \rangle + \frac{1}{4} \theta^2_y + \eta^2_y \right) 
  \nonumber \\
  & \mp \frac{\hbar^2}{m} \frac{\ep}{\ek} {\rm Re} X_{yy} \biggr] + (x \leftrightarrow y) + O(s^2),
%
%
\end{align}
%
Finally $\chi_{{\rm occ}2}$ becomes
\begin{equation}
   \chi_{\rm occ2} = \frac{e^2}{\hbar^2} \sum_{\pm,\k,\sigma} f(\pm E) \biggl[
	\mp \frac{\varepsilon}{2E} \{ \theta_x^2 \Delta_y \eta_y 
	- \theta_x \theta_y \Delta_x \eta_y + (x\leftrightarrow y)\}
   \pm \frac{\hbar^2}{2m} \frac{\varepsilon}{E} \sigma(\theta_x \eta_y - \theta_y \eta_x)
     \biggr] +O(s^2).
\end{equation}

In the total of these contributions, some terms cancel with each other. 
In the zeroth order of $s$ ($s^0$), there are terms proportional to 
$\frac{\hbar^2}{m}\langle y^2 \rangle$ and $\frac{\hbar^2}{m}\theta_x^2$ 
in $\chi_{\rm occ1}$ and $\chi_{\rm inter}$. 
However, the latter cancels with each other and only the former appearing in 
$\chi_{\rm occ1}$ contributes to the total susceptibility in the zeroth order.
In the previous paper,\cite{ogata3} we call this contribution as 
\lq\lq intraband atomic diamagnetism'', which is shown as 
$\chi_{\rm atomic}$ in Eq.~(\ref{eq:chi_atomic}).

Collecting the contributions proportional to $\hbar^2/m$ and $\langle y^2 \rangle$ 
and using integration by parts for terms in $\chi_{\rm FS}$, we obtain
\begin{align}
  &\chi_{2} = \frac{e^2}{\hbar^2} \sum_{\pm,\k,\sigma} f(\pm\ek) 
  \biggl[  \mp \frac{\rsq}{4\ek}
  (\varepsilon \varepsilon_{xx} + \Delta \Delta_{xx} -\varepsilon^2\thex^2) \nonumber \\
  &\mp \frac{\hbar^2}{4m} \left(\frac{s\varepsilon}{2Et} 
  (\varepsilon_{xx}-\varepsilon\theta_x^2) 
  - \frac{\varepsilon}{E} {\rm Re} X_{yy} \right)\biggr] + (x\leftrightarrow y) +O(s^2). 
\end{align}
where we have used the relations of Eqs.~(\ref{eq:SpRel0}) and (\ref{eq:SpRel4}).
Finally when we use the relation Eq.~(\ref{eq:AppCnew3}), we obtain Eq.~(\ref{eq:chi_2}).
The last term in $\chi_{\rm FS-P}$ is the usual Pauli paramagnetism, Eq.~(\ref{eq:chi_Pauli}).
The orbital-Zeeman (OZ) cross-terms are characterized by the presence of $\sigma$
and they appear in $\chi_{\rm FS}$, $\chi_{\rm FS-P}$, and $\chi_{\rm occ2}$. 
Their total becomes
\begin{align}
   \chi_{\rm OZ} &= \frac{e^2}{\hbar^2} \sum_{\pm,\k,\sigma} f'(\pm E) 
\frac{\hbar^2}{m} \sigma
\left[  (\varepsilon(\eta_x \theta_y -\eta_y \theta_x)-
\frac{\Delta}{4E} (E_x \theta_y - E_y \theta_x) \right] \nonumber \\
&+\frac{e^2}{\hbar^2} \sum_{\pm,\k,\sigma} f(\pm E) 
\biggl[ \pm \frac{\hbar^2}{2m} \sigma \frac{\varepsilon}{E} 
(\theta_x \eta_y - \theta_y \eta_x) \biggr] +O(s^2),
\end{align}
where we have used Eq.~(\ref{eq:SpRel2}) for $\Delta_x$ and $\Delta_y$.
Then using the integration by parts in the second term and using Eq.~(\ref{eq:SpRel2}) 
again, we obtain $\chi_{\rm OZ}$ in Eq.~(\ref{eq:chi_OZ}). 
The other terms lead to Eq.~(\ref{eq:chi_1}).

\section{$l'$ summation in $\chi_{\rm inter}$\label{sec:chiinter}}

To carry out the summation in $\chi_{\rm inter}$ Eq.~(\ref{eq:chiinter}), 
we first consider the case of $l'=\mp$. From (F2) and (F8), we have
\begin{align}
  M_{\pm \mp\sigma}^z = \mp \frac{e}{2\hbar} \Delta (\eta_x \theta_y - \eta_y \theta_x),
\end{align}
where we have used a relation Eq.~(\ref{eq:SpRel2}). Then, we have
\begin{align}
	&-2\sum_{\pm,\k,\sigma} \frac{f(\pm\ek)}{(\pm\ek) - (\mp\ek)} |M_{\pm\mp\sigma}^z|^2
	=-\frac{e^2}{2\hbar^2} \sum_{\pm,\k,\sigma}  f(\pm E ) \left [ \pm \frac{\dks^2}{2\ek}(\eta_x\they-\eta_y\thex)^2 \right ].
\end{align}

Next, we consider the case of $l'\neq\pm,\mp$.
In this case, using Eq.~(\ref{eq:dd-dd}) we can rewrite $M_{\pm l'\sigma}^z$ as
\begin{align}
  M_{\pm l'\sigma }^z =& -\frac{ie}{2\hbar} \biggl[
A_x
  (\pm \ek -E_{l'})  \int \frac{\partial \upms}{\partial k_y} u_{l'\k\sigma}\dr 
+B_x
  (\mp\ek - E_{l'}) \int \frac{\partial \umps}{\partial k_y} u_{l'\k\sigma} \dr \nonumber \\
&\qquad \pm E_y
  \int \frac{\partial \upms}{\partial k_x} u_{l'\k\sigma} \dr \biggr] 
  - (x \leftrightarrow y), 
\end{align}
where
\begin{align}
  &A_\mu = \mp i \frac{\dks}{2\ek}\them \pm \frac{s}{2t} \frac{\ep}{\ek}\epm \nonumber \\
  &B_\mu = -i\frac{\ep}{2\ek}\them \mp \eta_\mu 
	 \mp i\frac{s\varepsilon}{2t}\them - \frac{s}{2t}\frac{\dks}{\ek}\epm,
\end{align}
for $\mu=x, y$. Then, $M_{\pm l'\sigma}^z$ can be rewritten as
\begin{align}
  M_{\pm l'\sigma}^z = (M_1^{xy} + M_2^{xy}) - (x \leftrightarrow y),
\end{align}
with
\begin{align}
  &M_1^{\mu\nu} \equiv -\frac{ie}{2\hbar} 
	\{ A_\mu (\pm \ek - E_{l'}) \mp E_\mu\} \int \frac{\partial \upms}{\partial k_\nu} u_{l'\k\sigma} \dr \nonumber \\
  &M_2^{\mu\nu} \equiv -\frac{ie}{2\hbar}
	\{ B_\mu (\pm \ek - E_{l'}) \mp 2\ek B_\mu \} \int \frac{\partial \umps}{\partial k_\nu} u_{l'\k\sigma} \dr. 
\end{align}

Using these abbreviations, $|M_{\pm l'\sigma}^z|^2$ becomes
\begin{align}
  |M_{\pm l'\sigma}^z|^2 =& |M_1^{xy}|^2 + |M_2^{xy}|^2 + (M_1^{xy} M_2^{xy*} + {\rm c.c.})
  - M_1^{xy} M_1^{yx*} - M_2^{xy} M_2^{yx*} - (M_1^{xy} M_2^{yx*} + {\rm c.c.}) \nonumber \\
  &+(x \leftrightarrow y).
\end{align}
Thus, we need to calculate the six types of matrix elements.
In these calculations, we can write the $l'$ summation in a form,
\begin{align}
  \sum_{l'\neq \pm, \mp} \frac{(\pm \ek - E_{l'})^n}{\pm \ek -E_{l'}} \int X^* u_{l'\k\sigma} \dr \int u_{l'\k\sigma}^* Y\dr,
\end{align}
with $n=0,1,$ and $2$, and 
\begin{align}
  X,Y = \frac{\partial \upm}{\partial k_\mu}, \frac{\partial \ump}{\partial k_\nu},\quad {\rm etc.}
\end{align}
We can carry out these $l'$ summation in the following ways:\\
(a) $n=0$ case:\\
The denominator $\pm \ek-E_{l'}$ is in the zero-th order with respect to $s$, 
and the numerator is in the second order of $s$ because the prefactors 
$E_\mu$ and $EB_\mu$ are both in the order of $O(s)$.
Therefore, we can neglect this contribution of $n=0$ in the calculation 
in the order of $O(s)$.\\
(b) $n=1$ case:\\
Using the completeness condition, $\sum_{l'} u_{l'}(\r) u_{l'}^* (\r')=\delta(\r-\r')$,
we obtain
\begin{align}
  \sum_{l'\neq \pm,\mp} \int X^* u_{l'} \dr \int u_{l'}^* Y \dr 
  =\int X^*Y \dr - \int X^* \upm \dr \int \upms Y \dr - \int X^* \ump \dr \int \umps Y\dr.
	\label{eq:inter-n1}
\end{align}
(c) $n=2$ case:\\
Similarly, using the completeness condition, 
\begin{align}
  \sum_{l'\neq\pm,\mp} \frac{(\pm \ek - E_{l'})^2}{\pm \ek -E_{l'}} \int X^* u_{l'} \dr \int u_{l'}^* Y\dr
  &=\sum _{l' \neq \pm,\mp} \int X^* (\pm \ek -H_{\k\sigma})u_{l'}\dr \int u_{l'}^* Y \dr \nonumber \\
  &=\int X^* (\pm \ek - H_{\k\sigma}) Y \dr \mp 2\ek \int X^* \ump \dr \int \umps Y \dr.
	\label{eq:inter-n2}
\end{align}

Using the formulae (F1)-(F8), (\ref{dschroe}), (\ref{eq:inter-n1}), and (\ref{eq:inter-n2}), 
we can carry out the summation of all the combinations as follows,
\begin{align}
  &\sum_{l'\neq \pm,\mp} \frac{|M_1^{xy}|^2}{\pm \ek-E_{l'}} =
  \frac{e^2}{4\hbar^2}\biggl[ |A_x|^2 \left\{ -\frac{\hbar^2}{2m} \pm \frac{1}{2} E_{yy}
  \mp \frac{\varepsilon^2}{2\ek} \they^2 \mp 2E\eta_y^2 \right\}
  \mp 2E_x {\rm Re}[A_x] \langle y^2 \rangle + O(s^2) \biggr], \\
  &\sum_{l'\neq \pm,\mp} \frac{|M_2^{xy}|^2}{\pm \ek-E_{l'}}=
  \frac{e^2}{4\hbar^2} \biggl[ |B_x|^2 \left\{ -\frac{\hbar^2}{2m} \mp \frac{1}{2} E_{yy}
  \pm \frac{\varepsilon^2}{2\ek} \they^2 \pm 2E\eta_y^2  
  \mp 2\ek \langle y^2 \rangle \right\} + O(s^2) \biggr],\\
  &\sum_{l'\neq \pm,\mp} \frac{M_1^{xy}M_{2}^{xy*}}{\pm \ek-E_{l'}}=
  \frac{e^2}{4\hbar^2} \bigg[ A_xB_x^* 
  \left\{ \mp \frac{i}{2} (2\epy\they +\ep\theyy) - \frac{1}{2\ek}(\Delta\epyy - \varepsilon\Delta_{yy}-\varepsilon \Delta\they^2) \right\} + O(s^2) \biggr],\\
%
  &\sum_{l'\neq \pm,\mp} \frac{M_1^{xy}M^{yx*}_{1}}{\pm \ek-E_{l'}}= 
  \frac{e^2}{4\hbar^2} \biggl[ \pm A_x A_y^*
	\left( \frac{1}{2} E_{xy} - \frac{\ep^2}{2E} \thex\they -2E \eta_x \eta_y \right)  + O(s^2) \biggr],\\
  &\sum_{l'\neq \pm,\mp} \frac{M_2^{xy}M_2^{yx*}}{\pm \ek-E_{l'}}= 
	\frac{e^2}{4\hbar^2} \biggl[ \mp B_x B_y^*
  \left( \frac{1}{2} E_{xy} -\frac{\varepsilon^2}{2E} \thex\they -2E \eta_x \eta_y \right) + O(s^2) \biggr],\\
  &\sum_{l'\neq \pm,\mp} \frac{M_1^{xy}M_2^{yx*}}{\pm \ek-E_{l'}}= 
	\frac{e^2}{4\hbar^2} \biggl[ A_x B_y^*
  \left \{ \mp \frac{i}{2} (\epx\they + \epy\thex +\varepsilon \thexy) 
  - \frac{1}{2\ek}(\Delta\epxy -\varepsilon\Delta_{xy} - \varepsilon\Delta \thex\they) 
   \right\}+ O(s^2) \biggr].
\end{align}
Here we have used the relation in Eq.~(\ref{eq:SpRel3}). 
%
%
Then, substituting $A_\mu$ and $B_\mu$, and keeping the terms up to the order of 
$O(s)$, we obtain Eq.~(\ref{eq:chiInterFinal}).
\end{widetext}

\bibliography{main}
%

\end{document}